\documentclass[twocolumn,
 amsmath,amssymb,
 aps,
 pra,
 superscriptaddress,
]{revtex4-1}

\usepackage{graphicx}
\usepackage{dcolumn}
\usepackage{bm}
\usepackage{natbib}
\usepackage[utf8]{inputenc}
\usepackage{siunitx}
\usepackage{tabularx}
\usepackage{color}
\usepackage{transparent}
\graphicspath{{../Figures/}}
\usepackage{tikz}
\usepackage{tikz-timing}[2009/05/15]

\begin{document}

\title{Trapping, Shaping and Isolating of Ion Coulomb Crystals via State-selective Optical Potentials}

\author{Pascal Weckesser}
\affiliation{Albert-Ludwigs-Universit\"at Freiburg, Physikalisches Institut, Hermann-Herder-Straße 3, 79104 Freiburg, Germany}
 \author{Fabian Thielemann}
\affiliation{Albert-Ludwigs-Universit\"at Freiburg, Physikalisches Institut, Hermann-Herder-Straße 3, 79104 Freiburg, Germany}
 \author{Daniel Hoenig}
\affiliation{Albert-Ludwigs-Universit\"at Freiburg, Physikalisches Institut, Hermann-Herder-Straße 3, 79104 Freiburg, Germany}
 \author{Alexander Lambrecht}
\affiliation{Albert-Ludwigs-Universit\"at Freiburg, Physikalisches Institut, Hermann-Herder-Straße 3, 79104 Freiburg, Germany}
 \author{Leon Karpa}
\affiliation{Albert-Ludwigs-Universit\"at Freiburg, Physikalisches Institut, Hermann-Herder-Straße 3, 79104 Freiburg, Germany}
\affiliation{Leibniz University Hannover, Institute of Quantum Optics, Welfengarten 1, 30167 Hannover, Germany}
\author{Tobias Schaetz}%
\affiliation{Albert-Ludwigs-Universit\"at Freiburg, Physikalisches Institut, Hermann-Herder-Straße 3, 79104 Freiburg, Germany}

\date{\today}

\begin{abstract}

For conventional ion traps, the trapping potential is close to independent of the electronic state, providing confinement for ions dependent primarily on their charge-to-mass ratio $Q/m$.
In contrast, storing ions within an optical dipole trap results in state-dependent confinement.
Here we experimentally study optical dipole potentials for $^{138}\mathrm{Ba}^+$ ions stored within two distinctive traps operating at \SI{532}{\nano \meter} and \SI{1064}{\nano \meter}. 
We prepare the ions in either the $6\mathrm{S}_{\mathrm{1/2}}$ electronic ground or the $5\mathrm{D}_{\mathrm{3/2}}$/ $5\mathrm{D}_{\mathrm{5/2}}$ metastable excited state and probe the relative strength and polarity of the potential.
On the one hand, we apply our findings to selectively remove ions from a Coulomb crystal, despite all ions sharing the same $Q/m$.
On the other hand, we deterministically purify the trapping volume from parasitic ions in higher-energy orbits, resulting in reliable isolation of Coulomb crystals down to a single ion within a radio-frequency trap.


\end{abstract}

\pacs{Valid PACS appear here}

\maketitle

\section{\label{sec:level1} Introduction}

Conventional ion traps are based on a combination of static electric and magnetic fields (Penning traps) or radio-frequency (rf) fields (Paul traps) \cite{paul1990electromagnetic,dehmelt1991experiments,wineland2013nobel}, providing an established tool to isolate individual ions and store them for durations beyond hours and days.
The trapping potential is typically identical for ions of the same charge-to-mass ratio $Q/m$ and close to independent of their internal electronic state, with highly excited Rydberg ions being a recently reported exception \cite{higgins2017single}.
This unique feature allows to control the internal (electronic) and common external (motional) degrees of freedom, providing a basis for quantum information processing \cite{monroe2002quantum}, quantum simulation \cite{blatt2012quantum} and quantum metrology \cite{wineland2011quantum}.
For the latter, state-insensitive Paul traps enable state-of-the art atomic clocks \cite{ludlow2015optical} and the derivation of fundamental constants with their potential change in time \cite{safronova2014highly,kozlov2018highly}.
For these systems, there is the underlying assumption of proper isolation of the quantum system from its environment, dominated by black-body radiation \cite{safronova2013blackbody,beloy2014atomic} and residual stray electric fields \cite{huntemann2016single,beloy2018faraday}.

However, this assumption incorporates loading a dedicated number of ions into the trap, while prohibiting spurious contamination of the trapping volume by 'parasitic' ions.
These parasitic ions might be (i) of the same species, but on orbits of high energy; that is, hidden in the large trapping volume of the rf-trap, (ii) different isotopes or (iii) other ionic species or molecules.
Their presence can be difficult to detect directly by monitoring fluorescence.
They might lack geometric overlap with focused cooling and detection lasers, feature different electronic states and related transitions that remain far off-resonant or might lack a closed cycling transition.
Still, their residual long-range Coulomb interaction diminishes the isolation of the desired, closed quantum system.
In some experimental protocols parasitic ions might still be detected. As they remain subjected to sympathetic cooling by the Coulomb Crystal, they will eventually appear as additional lattice sites, evidenced as dark spots within the crystal’s lattice. 
Such events will occur randomly, potentially even minutes after the loading process has presumably ended. \cite{GuggemosHeinrichHerrera-SanchoEtAl2015}.
To remove parasitic ions of sufficiently deviating $Q/m$, methods based on the original concept of the quadrupole mass filtering have been refined.
In addition, resonant or parametric excitation is capable of increasing the kinetic energy of trapped ions beyond the potential depth \cite{drewsen2004nondestructive,nagerl1998coherent,razvi1998fractional,sudakov2000excitation,schmidt2020mass}.
However, it remains difficult to assure deterministic removal of parasitic ions in conventional ion-traps in an efficient way.
This especially holds true for ions with the same $Q/m$. 

Our approach, to isolate and prepare a given number of ions, builds on the recently demonstrated optical trapping of ions and Coulomb crystals in the absence of any rf fields \cite{schneider2010optical,huber2014far,lambrecht2017long,schaetz2017trapping,schmidt2018optical,karpa2019trapping}.
Even though the concepts behind rf and optical trapping are closely related \cite{cormick2011trapping}, there remain substantial differences and prospects.
Optical confinement can be achieved in an optical dipole trap (ODT), where the trap shrinks from the scale of $\sim10\%$ of the ion-electrode distance of the rf-trap to the waist of the laser beam.
The potential depth typically decreases by five orders of magnitude and the confinement predominantly depends on the electronic transition of choice and its ac Stark shift \cite{grimm2000optical}.
If the ODT in its simplest realization as a single focused Gaussian beam is red-detuned with respect to the transition, the ion experiences an attractive potential, whereas a blue-detuned laser acts repulsively.
Applying these state-dependent forces to ions confined in a common, largely state-insensitive rf-trap has been an integral part of quantum computation and simulation for the past decades.
For example, laser-beam intensity gradients can be applied to induce state-dependent displacements, implementing phase-gates \cite{blatt2008entangled,leibfried2003experimental} or mediating effective spin-spin interaction \cite{porras2004effective,friedenauer2008simulating}.
Furthermore, it has been shown that intracavity optical fields allow to localize ions to single lattice sites, while controlling the motional mode spectrum of 1D and 3D Coulomb crystal \cite{linnet2012pinning,karpa2013suppression,laupretre2019controlling,bylinskii2016observation}
These experiments are still operated with a continuously running rf-trap, where the dipole force modulates the 3D potential.

Here we investigate state-dependent optical potentials in the absence of rf fields.
We introduce a novel technique enabling individually addressable ion removal from a Coulomb crystal with shared $Q/m$, while further providing deterministic isolation from parasitic ions.
We realize this by intermittently transferring $^{138}\mathrm{Ba}^+$ ions from an rf-trap into two  distinct, far off-resonant single beam ODTs.
In a first step, we investigate the electronic state-dependent trap depths at a wavelength of \SI{1064}{\nano\meter} (NIR).
By changing to a blue-detuned dipole trap at \SI{532}{\nano\meter} (VIS), the dipole force acts repulsively for the 5D metastable manifolds, while providing confinement for the 6S electronic ground state.
In this way, by shelving the ion in the metastable state we can selectively remove dedicated ions from the ODT.
Returning into the rf-trap allows to continue working with a smaller Coulomb crystal.
In addition, we show how the intermediate transfer into an ODT further removes any parasitic ions on orbits of higher energy, isolating the remaining ions in the Coulomb crystal.

\section{Experimental Setup}
\begin{figure}[b!]
	\centering
  \includegraphics[width=\columnwidth]{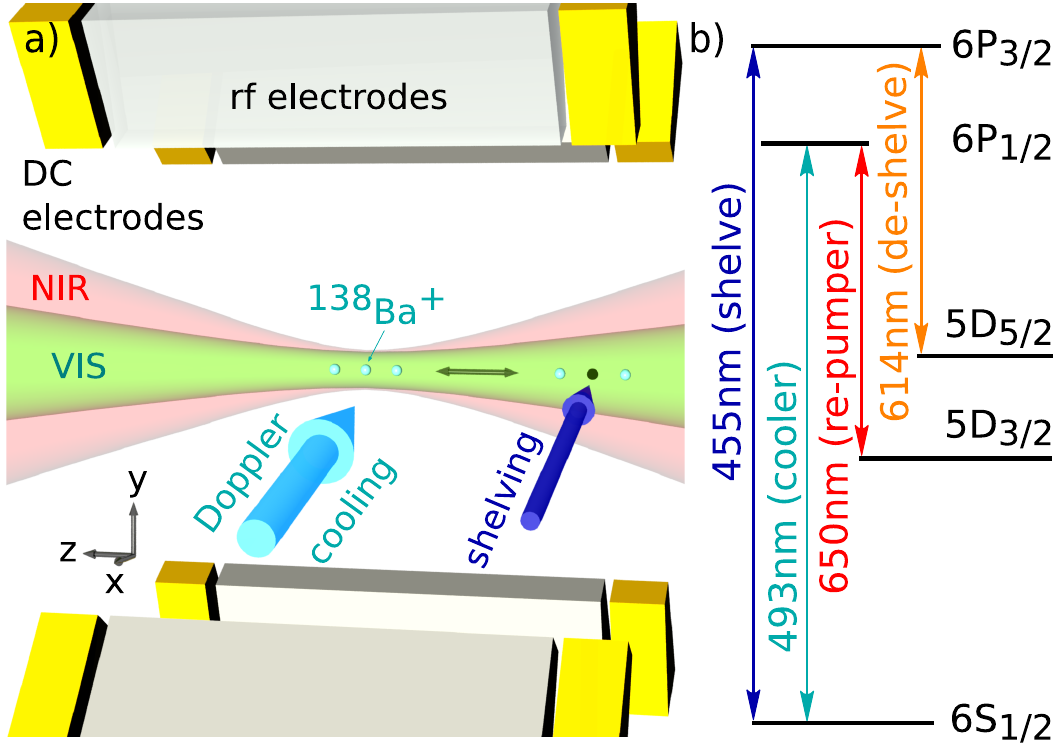}
	\caption{Overview of the experimental setup (not to scale) and the $^{138}\mathrm{Ba}^+$ level scheme.
a) Indicated are the Doppler cooling lasers (including the cooler and the repumper), both optical dipole traps (VIS and NIR) along the axial direction of the Paul trap, as well as the shelving laser being perpendicularly aligned to the ion Coulomb crystal.
dc electrodes are used for axial confinement, axial transport and stray-field compensation.
Further electrodes for radial stray-field compensation are not shown.
b) Simplified level scheme:
A total of four different lasers are used to control the internal state of the ion.
Doppler cooling is performed with the \SI{493}{\nano\meter} and \SI{650}{\nano\meter} lasers.
Additonal lasers at \SI{455}{\nano\meter} (\SI{614}{\nano\meter}) are used for shelving into (out of) the $5\mathrm{D}_{5/2}$ manifold.
}
	\label{fig:Setup}
\end{figure}

The experimental setup is an adapted version of a previously described apparatus used for optical trapping of ions \cite{lambrecht2017long,schmidt2018optical}. 
A schematic sketch of the ion trap and some of the relevant lasers are shown in Fig.~\ref{fig:Setup}.
Here, the linear segmented Paul trap is used for state-preparation and detection only.
The trap is located within an ultra-high vacuum chamber and is operated at a radio-frequency (rf) $\Omega_{\mathrm{rf}} \approx 2\pi \times$\SI{1.433}{\MHz} with a maximal zero-to-peak voltage $V_{\mathrm{rf}} \approx$\SI{1140}{\volt}.
With the ion-blade distance of $R_\mathrm{0} \approx$ \SI{9}{\milli\meter}, we typically achieve confinement in the radial (x and y) directions corresponding to secular frequencies of $\omega_{\mathrm{rad}} \approx 2\pi \times \{121, 123 \}$ \si{\kHz}.
The axial confinement is realized by dc electrodes at the end of the linear trap as illustrated in Fig.~\ref{fig:Setup}, with $\omega_{\mathrm{ax}} \approx 2\pi \times \{7.1, 8.5 \}$ \si{\kHz} for the presented experiments.

We load $^{138}\mathrm{Ba}^{+}$ ions by using a combination of laser ablation and subsequent two-photon ionization.
We irradiate a barium target with a pulsed laser beam operating at \SI{532}{\nano\meter} (pulse energy: $\leq$\SI{2}{\micro\joule}, repetition rate: $\leq$\SI{1}{kHz}, pulse duration: $<$\SI{1.2}{\nano\second}, waist on barium target: $24(4)$\,\si{\micro\meter}).
We choose these settings as they are below the threshold for direct ionization of Ba in our setup.
Otherwise this could cause several undesired effects, such as strongly increased target aging, stray charge deposition on the electrodes and loss of isotope-selectivity while loading.
We employ a random pointing technique by means of a scanning piezoelectric mirror for the ablation laser in order to mitigate local target aging, ultimately stabilizing the loading rate.
Following the ablation pulses, we ionize the neutral Ba vapor via a two-step process with laser beams operating at \SI{553}{\nano\meter} and \SI{405}{\nano\meter} (not shown in Fig.~\ref{fig:Setup}) \cite{leschhorn2012efficient}.

\begin{table}[b]
  \begin{center}
    \begin{tabular*}{\linewidth}{l @{\extracolsep{\fill}} c r}
    \hline
    \hline
      Laser & VIS  & NIR \\
      \hline
      $\lambda$ & \SI{532}{\nano\meter}   & \SI{1064}{\nano\meter} \\
       $w_{\mathrm{x}}$ & $4.5(2)$\,\si{\micro\meter} & $5.4(2)$\,\si{\micro\meter} \\
      $w_{\mathrm{y}}$ & $4.1(2)$\,\si{\micro\meter} & $5.4(2)$\,\si{\micro\meter}\\
      $z_{\mathrm{R}}$ &  $130-154$\,\si{\micro\meter} &  $86$\,\si{\micro\meter} \\
      $P_\mathrm{ODT}$ & $\leq$ $7.13(25)$\,\si{\watt} & $\leq 20.0(2)$\,\si{\watt}\\
      \hline
      $U_{\mathrm{opt}}(6\mathrm{S}_{\mathrm{1/2}},\pm 1/2)$ & $\leq$ $30.5\, \mathrm{mK}  \times \mathrm{k}_{\mathrm{B}}$ & $\leq$ $13.4\, \mathrm{mK}  \times \mathrm{k}_{\mathrm{B}}$\\
       	$\omega_{\mathrm{rad,opt}}(6\mathrm{S}_{\mathrm{1/2}},\pm 1/2)$ & $\leq$  $2\pi \times 95.5\, \mathrm{kHz} $  & $\leq$  $2\pi \times 52.8\, \mathrm{kHz} $ \\
  	$U_{\mathrm{opt}}(5\mathrm{D}_{\mathrm{3/2}},\pm 1/2)$ & - - & $\leq$ $6.35\, \mathrm{mK}  \times \mathrm{k}_{\mathrm{B}}$\\
  	       	$\omega_{\mathrm{rad,opt}}(5\mathrm{D}_{\mathrm{3/2}},\pm 1/2)$ & - -  &  $\leq$  $2\pi \times 36.3\, \mathrm{kHz} $ \\
	$U_{\mathrm{opt}}(5\mathrm{D}_{\mathrm{3/2}},\pm 3/2)$ & - - & $\leq$ $1.28\,  \mathrm{mK}  \times \mathrm{k}_{\mathrm{B}}$\\
	  	      $\omega_{\mathrm{rad,opt}}(5\mathrm{D}_{\mathrm{3/2}},\pm 3/2)$ & - -  & $\leq$  $2\pi \times 16.3\, \mathrm{kHz} $ \\
 	\hline
 	    $E_{\mathrm{S}}$ & $ \leq 5$\,\si[per-mode=symbol]{\milli\volt\per\meter}  & $ \leq 10$\,\si[per-mode=symbol]{\milli\volt\per\meter} \\
 	    $\omega_{\mathrm{ax}}/(2\pi)$ & $ 7.2(1) \, \mathrm{kHz}$ & $8.5(1)\, \mathrm{kHz}$ \\
 	   $\omega_{\mathrm{rad,DC}}^2/(2\pi)^2$ & $ -(6.5(1))^2 \, \mathrm{kHz}^2$ & $ -(8.5(1))^2\, \mathrm{kHz}^2$ \\
 	     $\Delta t_{\mathrm{opt}}$ & \SI{500}{\micro\second} & \SI{2000}{\micro\second} \\
      \hline
      \hline
    \end{tabular*}
  \end{center}
 \caption{Optical dipole trap parameters for the \SI{532}{\nano\meter} (VIS) and \SI{1064}{\nano\meter} (NIR) lasers.
Presented are wavelength ($\lambda$), $1/e^2$ beam waist in x- and y-direction ($w_{\mathrm{x}}$ and $w_{\mathrm{y}}$), Rayleigh-length ($z_{\mathrm{R}}$), laser power ($P_{\mathrm{ODT}}$),  optical trap depth ($U_{\mathrm{opt}}$) for the $6\mathrm{S}_{\mathrm{1/2}}$ and the $5\mathrm{D}_{\mathrm{3/2}}$ state assuming a single ion at the focus of the ODT ($\pi$-polarization), radial trapping frequency within the ODT assuming no dc curvatures and stray field ($\omega_{\mathrm{rad,opt}}$), detection resolution for radial stray electric fields ($E_{\mathrm{S}}$), axial confinement ($\omega_{\mathrm{ax}}$), the related radial de-confinement ($\omega_{\mathrm{rad,DC}}$) and the chosen optical trapping duration ($\Delta t_{\mathrm{opt}}$).
}
 \label{tab:odt_parameters}
\end{table}

The lasers used for cooling and shelving of the $^{138}\mathrm{Ba}^{+}$ in the rf-trap and a simplified scheme of the relevant electronic energy states are depicted in Fig.~\ref{fig:Setup}b).
The ions are Doppler cooled ($T_{\mathrm{D}}=$\,\SI{365}{\micro\kelvin}) with two lasers at \SI{493}{\nano\meter} (cooling) and \SI{650}{\nano\meter} (repumping), respectively.
While Doppler cooling, we detect the ions along two orthogonal directions ($\hat{x}$ and $\hat{z}$) of the linear Paul trap via fluorescence imaging with two charge-coupled device (CCD) cameras.
By using laser light at \SI{455}{\nano\meter}, we can shelve a predetermined arbitrary sub-set of ions within the Coulomb crystal into the metastable $5\mathrm{D}_{\mathrm{5/2}}$ manifold.
This laser is aligned perpendicularly to the Paul trap's z-axis shifted by \SI{300}{\micro\meter} from its center as indicated in Fig.~\ref{fig:Setup}a).
In this configuration we optically pump any target ion(s) by shuttling the linear ion string along the axial direction of the Paul trap by electric dc control-fields to the position of the laser beam \cite{naegerl1999laser}.
While selected ions are shelved, the other ones remain in the Doppler cooling cycle, providing sympathetic cooling for the Coulomb crystal.
Once optically pumped, the ion remains in the $5\mathrm{D}_{\mathrm{5/2}}$ manifold with a $1/e$-lifetime of \SI{31.2}{\second}.~\cite{auchter2014measurement}
By using an additional laser at \SI{614}{\nano\meter}, we can depopulate the metastable manifold and the ion returns into the Doppler cooling cycle.

We can further transfer and confine the ions in two distinctive ODTs.
Similar as in Ref.~\cite{lambrecht2017long,schmidt2018optical}, the ODTs operate at either \SI{532}{\nano\meter} (VIS) or \SI{1064}{\nano\meter} (NIR), with Table~\ref{tab:odt_parameters} summarizing some of their important parameters.
Both ODTs are aligned along the axial direction of the Paul trap, entering the chamber from opposite directions.
In this configuration, each ODT confines the ions along the radial direction, whereas the axial confinement is still predominantly provided by dc control-fields.
This is necessary as the comparatively large Rayleigh length of the ODTs does not provide a significant axial confinement.
It is worth mentioning, that unlike in a conventional rf-trap, an ion confined in a Gaussian beam experiences position-dependent radial confinement $\omega_{\mathrm{rad,opt}}(z)$, as the laser beam diverges with increasing distance from the focal plane.

\section{Experimental Methods - Optical Ion Trapping and Thermometry}
\label{section:methods}

\begin{figure}[b]
	\centering
  \includegraphics[width=\columnwidth]{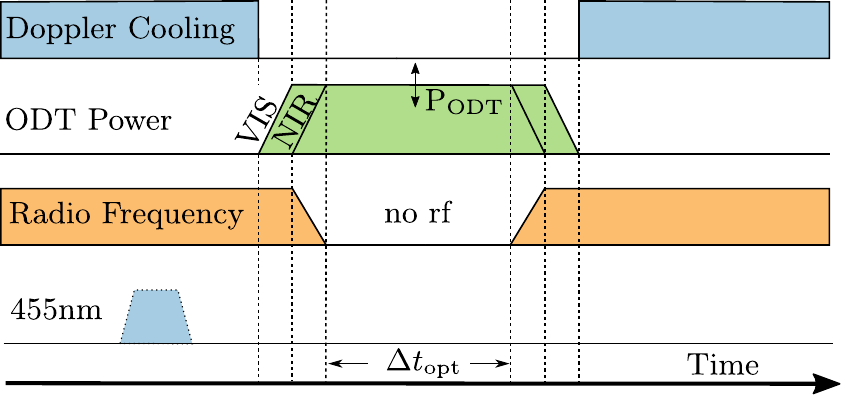}
	\caption{Protocol for intermediate ion trapping within an ODT in the absence of rf fields.
The presented protocol (not to scale) can be divided in three stages: (1) Preparation in the rf-trap: we prepare 1-4 ions close to Doppler temperature.
During this stage we can optionally pump dedicated ions into the metastable $5\mathrm{D}_{\mathrm{5/2}}$ manifold by illuminating a target ion with \SI{455}{\nano\meter}.
(2) Optical Trapping: the ions are transferred into the ODT.
We turn off the rf fields, while the electrostatic axial potential remains unchanged. 
(3) Detection in the rf-trap: we switch on the rf-fields again and detect the ions while Doppler cooling.
}
\label{fig:protocol}
\end{figure}

\textbf{Optical Trapping Protocol:} The optical trapping protocol is shown in Fig.~\ref{fig:protocol}.
Here we distinguish between three major phases: state-preparation, optical trapping and detection.
First, we prepare a linear Coulomb crystal of up to four ions close to the Doppler limit in the rf-trap.
While Doppler cooling, we can shelve individual ions of our choice into the metastable $5\mathrm{D_\mathrm{5/2}}$ manifold.
The remaining bright ions are optically pumped to the $6\mathrm{S}_{1/2}$ ($5\mathrm{D}_{3/2}$) state, by first turning off the \SI{493}{\nano\meter} (\SI{650}{\nano\meter}) laser light.
Once the ions are prepared, we gradually increase the power of the VIS ODT over \SI{100}{\micro\second} and subsequently decrease the rf-fields within \SI{100}{\micro\second} (for the NIR case both ramps occur simultaneously).
Applying these ramps allows us to transfer the ions into the ODT, without noticeable heating \cite{lambrecht2017long}.
It is important, that the centers of the rf trap and the ODTs are aligned with an accuracy of $<$\,\SI{500}{\nano\meter}.
Without the presence of an ODT, a single $^{138}\mathrm{Ba}^+$ ion requires on average $200(30)$\,\si{\micro\second} to escape from the trapping region of the rf-trap once the rf-fields are turned off.
To ensure negligible re-trapping effects, we choose a trapping duration of at least $\Delta t_{opt} =$\,\SI{500}{\micro\second}, which is significantly larger than the escape time.
In principle this duration can be chosen as large as several seconds \citep{lambrecht2017long}.
Afterwards, we switch on the rf fields and turn off the dipole trap.
Finally, we detect and count the remaining ions via fluorescence imaging.

\textbf{Electro-Optical Trap Depth $U_0$:} The optical trap depths $U_{\mathrm{opt}}$ in Table~\ref{tab:odt_parameters} represent the ac Stark shifts \citep{grimm2000optical} at the focus of the ODT.
Unlike neutral atoms, ions experience Coulomb forces and are consequently much more sensitive to electric fields.
Thus even for a single ion, one also needs to consider the contributions of radial dc curvatures $m \omega_{rad,DC}^2$ and stray electric fields $E_S$.
The former arise due to dc axial confinement, inevitably leading to defocussing in at least one radial direction \cite{maxwell1873treatise}.
Adding all contributions results in the smallest maximum along the direction of dc defocussing.
The difference between this local maximum and the absolute minimum is then the effective electro-optical trap depth $U_{0}$, with $U_{0} < U_{\mathrm{opt}}$ \citep{karpa2019trapping}.

We obtain $U_{0}$ and its uncertainty by bootstrapping our experimental values provided in Table~\ref{tab:odt_parameters}.
This is necessary, as we only detect an upper bound for the residual stray-electric fields $E_{\mathrm{S}}$.
The precise magnitude and direction of the stray field remains unknown for a single experimental realization.
However, as we are probing ensemble averages, we assume the residual stray fields to follow a random distribution.
For the presented analysis, we sample $E_{\mathrm{S}}$ following a Gaussian distribution with sigma being equal to our detection limits given in Table~\ref{tab:odt_parameters}.
We further sample the uncertainties of the waists ($w_x$, $w_y$) and dc curvatures.
As an example, calculating the NIR trap for maximal $P_{\mathrm{ODT}}$ by bootstrapping leads to $U_{0}^{NIR}(\mathrm{6S}_{\mathrm{1/2}}) = 11.1(1.0) \, \mathrm{mK} \times k_{B}$, which is smaller than $U_{\mathrm{opt}}^{NIR}(\mathrm{6S}_{\mathrm{1/2}}) = 13.4 \, \mathrm{mK} \times k_{B}$.

We extend $U_{0}$ to several ions by including the mutual Coulomb interaction while simultaneously considering the position-dependent ac Stark shift of the ODT.
For the Coulomb interaction, we perform a Taylor expansion of the electrostatic potential $\Phi_j(z_j) = \sum_{i \neq j} \frac{q^2}{4\pi\epsilon_{0}} \frac{1}{\|z_i - z_j\|}$ around the equilibrium position of the $\mathrm{j}^{\mathrm{th}}$ ion \citep{james1998quantum} along the radial direction $\hat{x}$.
The second order term amounts to the repulsive Coulomb curvature $m\omega_{C}(z_j)^2$, with typical values of several \si{\kHz} for $\omega_{C}/(2\pi)$ in our setup.
The resulting position-dependent, defocusing dc curvature is then calculated by $m\omega^2(z_j) = m(\omega_{rad,DC}(z_j)^2 + \omega_{C,j}^2)$.

\textbf{Thermometry:} Trapping ions optically allows to determine the ensemble average of the ions' kinetic energy and therefore temperature $T_{\mathrm{ion}}$ \cite{Schneider2012influence}.
Here we repeatably probe the trapping probability $p_{\mathrm{opt}}$ of a single ion for varying $U_{0}$.
This reveals the underlying temperature, as $p_{\mathrm{opt}}$ can be related to $T_{\mathrm{ion}}$ by the radial cutoff-model $p_{\mathrm{opt}}=1-e^{-2\xi}-2\xi e^{-\xi}$, with $\xi=U_{0}/T_{\mathrm{ion}}$ \cite{Schneider2012influence}.
This model assumes that each ion undergoes several oscillations within the ODT, allowing the ion to properly probe the potential.
For the presented analysis, the radial trapping frequencies are in the range of $2\pi \times [12, 48 ]$\,\si{\kHz}.
Choosing a trapping duration of $\Delta t_{\mathrm{opt}} =$\,\SI{2}{\milli\second} therefore fulfills the constraint.
The presented thermometry will only be used for the NIR ODT.
Here, the off-resonant scattering rate \cite{grimm2000optical} at maximal $P_{\mathrm{ODT}}$ is negligible ($\Gamma(\mathrm{6S_{1/2}}) \approx$\,\SI{10}{\Hz} and $\Gamma(\mathrm{5D_{3/2}}) \approx$\,\SI{19}{\Hz}) compared to the inverse trapping duration.
Therefore proper state-preparation is ensured with fidelity close to unity while measuring $T_{\mathrm{ion}}$. 

\section{State-selective potentials in the NIR trap}

\begin{figure}[b!]
	\centering
 \includegraphics[width=1.\columnwidth]{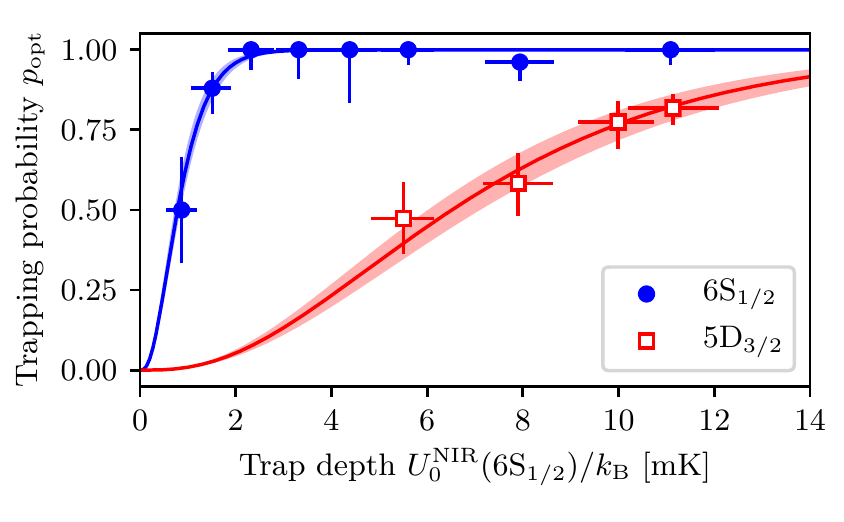}
	\caption{Demonstration and probing state-dependent averaged trap depths in the NIR ODT.
Illustrated are the trapping probability for a single ion depending on NIR laser intensity expressed as $U_{\mathrm{0}}^{\mathrm{NIR}}(6\mathrm{S}_{\mathrm{1/2}})$.
Full Circles: Experimental data of ions being prepared in the $\mathrm{6S}_{\mathrm{1/2}}$ state.
Fitting a radial cutoff model \cite{Schneider2012influence} with orthogonal distance regression (upper solid line) results in a temperature of $T_{\mathrm{ion}}(6\mathrm{S}_{\mathrm{1/2}}) = 355(29)$\,\si{\micro\kelvin}.
Open Squares: Data for an ion being prepared in the $\mathrm{5D}_{\mathrm{3/2}}$ state.
The lower solid line results by fitting the data with a modified radial-cutoff model $p_{\mathrm{opt}}=1-e^{-2\kappa_{\mathrm{0}}\xi}-2\kappa_{\mathrm{0}}\xi e^{-\kappa_{\mathrm{0}}\xi}$, while keeping the temperature fixed to the value obtained from the $6\mathrm{S}_{\mathrm{1/2}}$ evaluation.
Bootstrapping the uncertainty of the temperature allows to extract an upper 1$\sigma$ uncertainty for $\kappa_0$, yielding $\kappa_{\mathrm{0}} = 0.12(1)$.
Shaded regions: bounds corresponding to the fit standard errors.
Error bars: (trap depth) upper bounds of 1$\sigma$ uncertainty extracted from bootstrapping our experimental uncertainties; ($p_{\mathrm{opt}}$) upper bounds of 1$\sigma$ confidence intervals calculated from the underlying binomial distribution.
}
	\label{fig:nir_trap}
\end{figure}

Unlike in a rf trap, the confinement provided by an optical dipole trap is state-dependent.
Here, the ac Stark shift depends on the laser's polarization, the magnetic quantization axes and the Zeeman magnetic sublevel $m_{\mathrm{F}}$ within a manifold.

In the following, we probe state-dependent trap depths for a single $^{138}\mathrm{Ba}^+$ ion within the NIR trap.
We measure the state-dependent optical trapping probability for variable laser power and evaluate the results with the radial-cutoff model.
As discussed before, the model only depends on $U_{0}$ and $T_{\mathrm{ion}}$.
Knowledge of one quantity therefore allows the derivation of the other and vice versa.

For the presented analysis, we examine the $6\mathrm{S}_{\mathrm{1/2}}$ and $5\mathrm{D}_{\mathrm{3/2}}$ manifold.
It is important to mention that we examine averaged potentials, as our current state preparation lacks access to individual $m_{\mathrm{j}}$-levels (note that the nuclear spin equals $I = 0$. Therefore $m_{\mathrm{F}} = m_{\mathrm{j}}$).
For the $6\mathrm{S}_{\mathrm{1/2}}$ manifold this is insignificant, as both $m_\mathrm{j}$-levels experience the same ac Stark shift for an ODT with $\pi$-polarization.
We can therefore calculate $U_{0}^{\mathrm{NIR}}(6\mathrm{S}_{\mathrm{1/2}})$ as in previous investigations \cite{lambrecht2017long,karpa2019trapping}, whereas $U_{\mathrm{0}}^{\mathrm{NIR}}(5\mathrm{D}_{\mathrm{3/2}})$ needs to be evaluated experimentally.
We obtain the averaged trap depth by applying the radial cutoff-model twice.
We first use the known potential $U_{0}^{\mathrm{NIR}}(6\mathrm{S}_{\mathrm{1/2}})$ to derive a temperature.
We then use $T_{\mathrm{ion}}$ to obtain $U_{\mathrm{0}}^{\mathrm{NIR}}(5\mathrm{D}_{\mathrm{3/2}})$.

The experimental results of the trapping probability with respect to the laser's intensity are presented in Fig.~\ref{fig:nir_trap}.
For better comparison, the intensity is expressed in units of $U_{0}^{\mathrm{NIR}}(6\mathrm{S}_{\mathrm{1/2}})$, revealing the difference in state-dependent trap depth. 
Fitting the $6\mathrm{S}_{\mathrm{1/2}}$ data using the radial-cutoff model and the known $U_{0}^{\mathrm{NIR}}(6\mathrm{S}_{\mathrm{1/2}})$ yields $T_{\mathrm{ion}}(6\mathrm{S}_{\mathrm{1/2}}) = 355(29)$\,\si{\micro\kelvin}, consistent with the Doppler temperature of $T_{\mathrm{D}} =$\,\SI{365}{\micro\kelvin}.

Preparing the ion in the $5\mathrm{D}_{\mathrm{3/2}}$ manifold in our current scheme involves the scattering of additional photons compared to preparation in the $6\mathrm{S}_{\mathrm{1/2}}$ state.
However, the corresponding change of the ion's energy stemming from photon recoil amounts to just a few \si{\micro\kelvin}.
As this energy scale remains negligible compared to $T_\mathrm{D}$, we assume $T_{\mathrm{ion}}(5\mathrm{D}_{\mathrm{3/2}}) \approx T_{\mathrm{ion}}(6\mathrm{S}_{\mathrm{1/2}})$.

We derive $U_{\mathrm{0}}^{\mathrm{NIR}} (5\mathrm{D}_{\mathrm{3/2}})$ by introducing the ratio of the effective trap depths as a scaling factor $\kappa_{\mathrm{0}} = U_{\mathrm{0}}^{\mathrm{NIR}} (5\mathrm{D}_{\mathrm{3/2}}) / U_{\mathrm{0}}^{\mathrm{NIR}}(6\mathrm{S}_{\mathrm{1/2}})$.
We fit the $5\mathrm{D}_{\mathrm{3/2}}$ data with a modified radial-cutoff model $p_{\mathrm{opt}}=1-e^{-2\kappa_{\mathrm{0}}\xi}-2\kappa_{\mathrm{0}}\xi e^{-\kappa_{\mathrm{0}}\xi}$.
Assuming an unchanged $T_{\mathrm{ion}}$ while having $\kappa_{\mathrm{0}}$ as the only free parameter yields $\kappa_0 = 0.12(1)$.
For our setup, an ion prepared in the $6\mathrm{S}_{\mathrm{1/2}}$ state therefore experiences a trap depth $\sim 8.3(7)$ times larger compared to the ion being prepared in the $5\mathrm{D}_{\mathrm{3/2}}$ manifold.

Finally, we compare our findings with theory \cite{kaur2015magic}.
Assuming an equal occupation of all $m_j$-levels in the $5\mathrm{D}_{\mathrm{3/2}}$ manifold, allows the derivation of a theoretical ratio of the ac Stark shifts $\kappa_{\mathrm{opt}}^{\mathrm{theo}} = U_{\mathrm{opt}}^{\mathrm{NIR}} (5\mathrm{D}_{\mathrm{3/2}}) / U_{\mathrm{opt}}^{\mathrm{NIR}}(6\mathrm{S}) = 0.28$.
The difference between $\kappa_0$ and $\kappa_{\mathrm{opt}}^{\mathrm{theo}}$ is due to the fact, that the experimental result $\kappa_0$ includes the contributions of stray-electric fields and dc curvatures.
However, we can still deduce $\kappa_{\mathrm{opt}}^{\mathrm{exp}}$ numerically, by evaluating for which value our data set fulfills the condition $T(5\mathrm{D}_{\mathrm{3/2}}) \approx T(6\mathrm{S}_{\mathrm{1/2}})$.
We find $\kappa_{\mathrm{opt}}^{\mathrm{exp}} = 0.24(1)$.
Our experimental finding is in qualitative agreement with theory.
The remaining discrepancy could be explained by a non-uniform distribution of the $m_j$-levels, by contributions of $\sigma^-/\sigma^+$-polarization or by the residual displacement between the rf node and the NIR center.

\section{State-selective ion removal in the VIS trap}

In a next step we probe the state-dependent trapping performance for an ODT operated at \SI{532}{\nano\meter} (VIS).
Unlike in the NIR case, the VIS laser is blue-detuned for ions prepared in both the $5\mathrm{D}_\mathrm{3/2}$ and $5\mathrm{D}_\mathrm{5/2}$ manifold and thus acts repulsively, while remaining red-detuned for the $6\mathrm{S}_{\mathrm{1/2}}$ state.

\begin{figure}[b!]
	\centering
\includegraphics[width=\columnwidth]{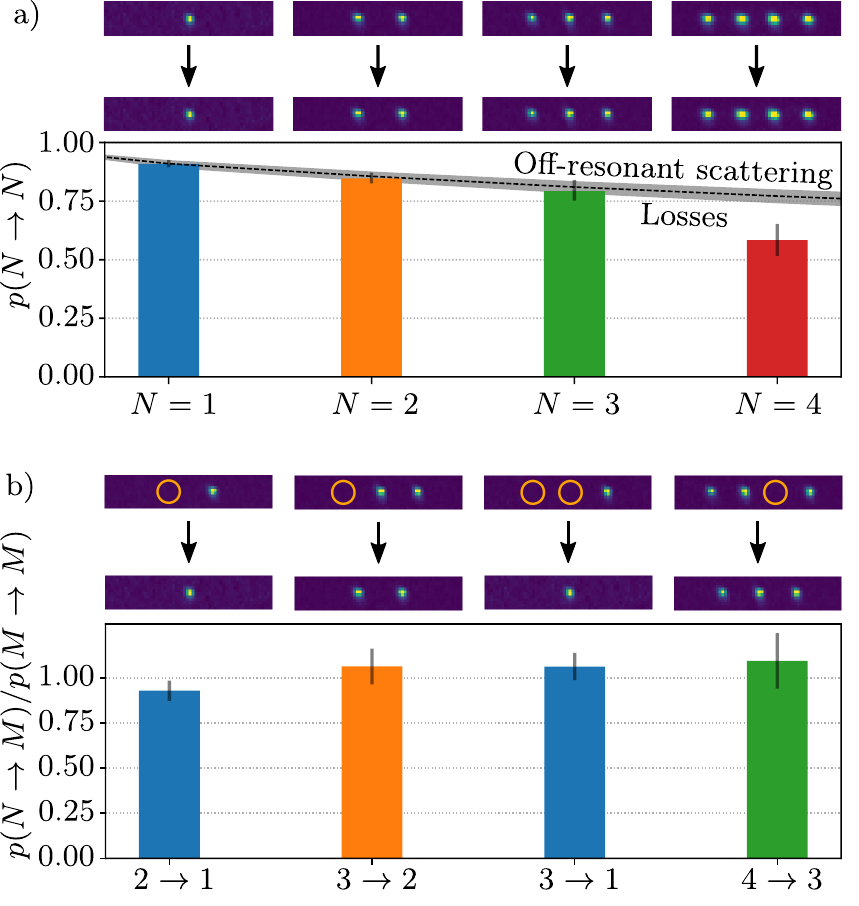}
	\caption{Demonstration of state-dependent trapping with the VIS ODT.
a) Optical trapping efficiency for $N= \{1,2,3,4\}$ ions being prepared in the $6\mathrm{S}_{\mathrm{1/2}}$ state.
Above: exemplary fluorescence images of ion Coulomb crystals before and after optical trapping are shown.
Below: corresponding measured optical trapping probability.
The error-bars denote the statistical uncertainty given by the calculated $1\sigma$ Wilson score interval.
The gray dashed line indicates the theoretical prediction of the loss rate by off-resonant scattering into the metastable D manifolds.
The gray dashed area represents the bounds of the theoretical prediction based on our experimental uncertainties.  
b) Performance of state-selective removal normalized to the optical trapping efficiency. 
Above: exemplary fluorescence images of ion Coulomb crystals before and after state-selective removal.
The orange circles within the fluorescence images mark the dark ions prepared in $5 \mathrm{D}_{\mathrm{5/2}}$ manifold.
Experimental results are shown below.
The color code of the bars equals the color code of the final ion number $M$ in a).
The measured values were normalized to the overall trapping performance of the final crystal size $p(M\rightarrow M)$, to obtain the efficiency of state-selective removal only.
Therefore some values are larger than $1$, but consistent with unity within their respective uncertainties.
}
	\label{fig:fidelity}
\end{figure}

To begin, we characterize the trapping performance for the $6\mathrm{S}_{\mathrm{1/2}}$ state for Coulomb crystals of up to four ions.
Once the ions are prepared, we transfer the crystal into the VIS ODT, being operated at maximum $P_{\mathrm{ODT}}$.
The outcome of these measurements and crysal size dependent trapping performance can be seen in Fig.~\ref{fig:fidelity}a).

As in a previous investigation~\cite{lambrecht2017long}, the performance of the VIS ODT is mainly limited by off-resonant scattering into the D manifolds.
We compare our experimental results with a simple rate equation model calculating the off-resonant scattering events into the D manifolds, being represented by the gray shaded area in Fig.~\ref{fig:fidelity}a).
Here for each linear Coulomb crystal configuration ($N=1 \dots 4$), the ions' positions within the ODT are calculated as in \cite{james1998quantum}.
Then we determine the local VIS intensity for each ion, assuming a gaussian beam propagation with the values provided in Table.~\ref{tab:odt_parameters}.
Subsequently, we compute the off-resonant scattering rate for each ion as in~\citep{grimm2000optical}, while further considering the branching ratio of the excited states ($P\to S$ vs $P \to D$) of $\sim 3:1$ \cite{de2015precision,arnold2019measurements}.

The result of this simple model is consistent with our observation for the case of up to three ions, whereas for the case of $N=4$ the measured $p_{\mathrm{opt}}$ is significantly lower, indicating that additional effects have to be considered.
For $N=4$ the length of the Coulomb crystal amounts to $\|z_1 -z_4 \| \approx$\,\SI{229}{\micro\meter}.
For the outermost ion the waist of the ODT expands to $\sim$\,\SI{6.2}{\micro\meter}.
Further including the mutual Coulomb interaction ($\omega^2_{C,j=2,3}/(2\pi)^2 = -(11.3)^2 \,\mathrm{kHz}^2$ and $\omega^2_{C,j=1,4}/(2\pi)^2 = -(7.9)^2 \,\mathrm{kHz}^2$), the outermost ion experiences the trap depth $U_{0,j=1,4}^{VIS}(\mathrm{6S}_{\mathrm{1/2}}) = 11.6(8) \, \mathrm{mK} \times k_{B}$.
In principle this trap depth should still be sufficient to reliably trap four ions optically.
However, these estimations assume ideal alignment between the wave vector of the VIS laser and the axis of the Paul trap.
A possible dislocation between the nodes displaces the ions during the transfer into the ODT, effectively heating the Coulomb crystal.
A similar effect has been observed in previous work \citep{schmidt2018optical}.
It should be noted that we compensate stray electric fields with a single ion at the ODT's focus.
Therefore the outermost ions might experience larger stray fields.

In the following we extend our analysis to the metastable $5\mathrm{D}_{\mathrm{5/2}}$ level.
By shelving selected ions in the $5\mathrm{D}_{\mathrm{5/2}}$ manifold and subsequently transferring the crystal into the VIS ODT, we selectively render the effective potential repulsive for the marked ions. 
This enables us to controllably remove any ion from the Coulomb crystal, even if all ions share the same $Q/m$.
We call this process \textit{state-selective removal}.

The experimental results on state-selective removal can be seen in Fig.~\ref{fig:fidelity}b).
Again we prepare Coulomb crystals up to four $^{138}\mathrm{Ba}^+$ ions, but now selectively shelve either one or two ions at different positions in the metastable $5\mathrm{D}_{\mathrm{5/2}}$ manifold.
Since the cooling lasers are off-resonant with respect to the shelved ions, they appear dark.
However, the Coulomb interaction still reveals their location within the Coulomb crystal, being indicated by the orange circles in Fig.~\ref{fig:fidelity}b).
For each configuration we obtain a removal efficiency $p(N \rightarrow M)$, with $N$ and $M$ being the initial and final ion number of choice, respectively.
In order to determine the efficiency of state-selective removal only, we need to normalize our measured values by the trapping performance of the final Coulomb crystal configuration, leading to $p(N\rightarrow M)/p(M\rightarrow M)$.
We find consistency with unity.

Currently, the success rate of state-selective removal is not limited by the efficiency of the removal process itself, but rather by the optical trapping performance of the VIS ODT (see Fig.~\ref{fig:fidelity}a).
This could be improved by lowering $P_{\mathrm{ODT}}^{\mathrm{VIS}}$, effectively reducing losses by off-resonant scattering, or by including repumpers at \SI{614}{\nano\meter} and \SI{650}{\nano\meter} \citep{lambrecht2017long}, once the ions originally prepared in the $5\mathrm{D}_{\mathrm{5/2}}$ manifold left the ODT.
Nevertheless, the process of state-selective removal can still be observed with high fidelity, by repeating the protocol until a successful event occurs.
As a showcase, by performing state-selective removal for the case of $3\rightarrow 2$ ions with our current efficiencies, we obtain a theoretical success rate of $\sim 99.7\%$ after three repetitions.

\section{Isolating a single ion}

In the last section, we apply \textit{state-selective removal} to build an efficient deterministic single ion source for rf-traps, while further ensuring isolation from any parasitic ions.

During the course of ion loading, random number of atoms are ionized.
Loading a single ion in the comparatively large trapping volume of a linear Paul trap can therefore be challenging.
In particular, the ions are created at random positions in the rf-trap, most probably starting on a higher energy trajectory where $E_{\mathrm{kin}} = e\cdot U_{\mathrm{rf}}(\vec{r})$.
These ions exhibit a long capture time into the Coulomb crystal, which can take up to several minutes \citep{GuggemosHeinrichHerrera-SanchoEtAl2015a}.

There are different approaches loading only a single ion.
For the case of laser ablation, one can reduce the power of the ablation laser.
While creation and capture of more than one ion can be largely suppressed, it drastically reduces the loading efficiency \cite{hendricks2007all}.
A rather fast alternative is to first load a larger ion Coulomb crystal and subsequently induce loss of ions by pulsing the rf and dc confinement \cite{schnitzler2010focusing}.
Both techniques however, can not guarantee the absence of parasitic ions, following higher energy trajectories.

\begin{figure}[b!]
	\centering
 \includegraphics[width=\columnwidth]{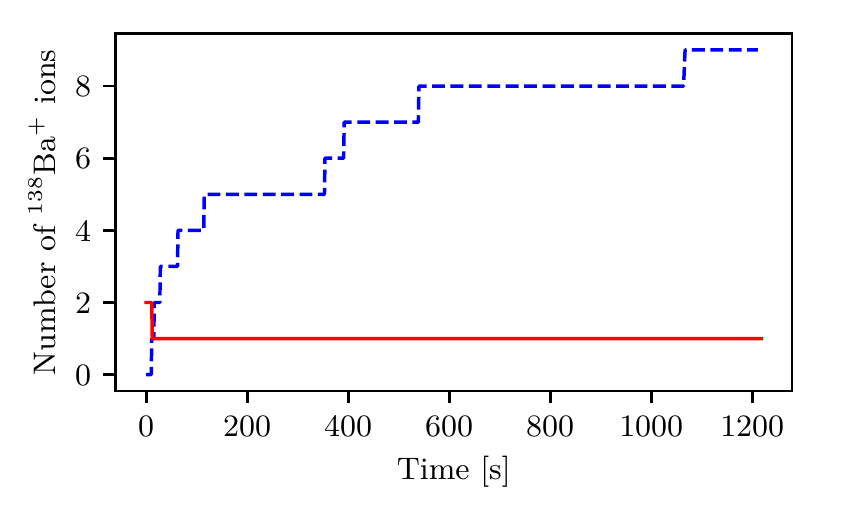}
	\caption{Efficient loading and isolation of a single ion in an rf-trap by cutting-off parasitic ions using an intermediate step of optical trapping in the VIS ODT.
Dashed blue curve: 
We apply ablation loading and photo-ionization in the first \SI{3}{\second} of the experiment and without further ablation nor photo-ionization track the number of bright $^{138}\mathrm{Ba}^+$ ions over \SI{20}{\minute}.
Even after \SI{20}{\minute} the size of the Coulomb crystal still increases, revealing the presence of former hidden ions within the rf-trap.
Solid red curve: 
We apply ablation loading for \SI{3}{\second}.
Unlike before, once two ions form a Coulomb crystal, we immediately remove one ion by state-selective removal.
The shallow depth and the small volume of the VIS ODT allows the storage of a finite number of ions, while parasitic ions on higher energy trajectories are effectively removed from the former trapping volume of the rf-trap.
Transferring the ions from the ODT into the rf-trap again can be understood as deterministic loading and purification of the rf-trap.
}
\label{fig:deterministic}
\end{figure}

Optical trapping of ions in absence of any rf-fields overcomes this difficulty.
It provides effective isolation of the ion ensemble of dedicated size down to a single ion, while allowing high repetition rates.
The shallow optical trap and related microscopic fraction of the trapping volume confines a well defined maximal number of ions.
In our experimental setup, the trapping volume is reduced by $\sim 9$ orders of magnitude.
Due to the reduction of trapping volume, all parasitic ions on higher energy trajectories are effectively ejected.
The remaining size of the purified Coulomb crystal within the ODT can then be shaped to the desired size by state-selective removal.
In the following, as a showcase, we demonstrate the effective isolation of a single ion, after being retransferred in the rf-trap.

We start by characterizing the capture rate of ions from higher energy trajectories into the Coulomb crystal by Doppler cooling.
We apply ablation loading while simultaneously turning on the photo-ionization and Doppler cooling lasers for \SI{3}{\second}.
Without further ablation and photo-ionization, we track the number of bright ions in the rf-trap for the subsequent \SI{20}{\minute}.
The result of an exemplary experimental run can be seen as the blue dashed curve in~Fig.~\ref{fig:deterministic}.
Most ions are created on higher orbit-trajectories and therefore require several minutes before they appear bright in the center of the rf-trap.
Even after \SI{20}{\minute}, the absolute number of ions within the Coulomb crystal still increases.
This dynamic reveals the difficulty to deterministically prepare and isolate N ions within an rf-trap. 

In a second measurement, shown as the red solid curve in~Fig.~\ref{fig:deterministic}, we extend the loading protocol by our state-selective removal scheme.
As an illustration, we start with two bright ions.
We immediately remove one, reducing the crystal size down to a single ion.
Afterwards, we observe that over the course of \SI{20}{\minute} no other ions appear in the rf-trap.
We see this as evidence for removing any parasitic ions via the intermittent transfer into the shallow VIS ODT.
The presented scheme provides an efficient and deterministic method for loading and isolating a single ion.
In principle this scheme can be extended to larger number of ions.

\section{Conclusion and outlook}

In summary, we demonstrated state-dependent confinement for Coulomb crystals in two independent single-beam optical dipole traps.
Exploiting these techniques allows us to create state-dependent potentials or to reliably remove a given ion from a Coulomb crystal, despite all ions sharing the same $Q/m$.
We further showed that the intermittent transfer of a Coulomb crystal into an ODT allows reliable and fast isolation down to a single ion for any final trapping scheme, such as Paul and Penning traps.
The presented work opens the door for numerous applications, some of which will be discussed in the following.

The newly presented scheme of state-selective removal is directly applicable for a number of of earth-alkaline ions (e.g. $\mathrm{Ca}^+$, $\mathrm{Sr}^+$, $\mathrm{Ba}^+$ and $\mathrm{Yb}^+$), which all feature metastable electronic states between the ground and first excited state.
We note that in principle this scheme can be adapted for ions without such metastable states (such as $\mathrm{Be}^+$ and $\mathrm{Mg}^+$), by individually exposing the target ions with additional blue-detuned optical light fields focused on the selected positions while the entire Coulomb crystal is confined in an attractive ODT.
Since the target ions don't have to be kept in the regime of low off-resonant scattering induced by the blue-detuned laser, this can be accomplished with near-resonant laser beams operated at moderate optical powers.

Removing a target ion from a Coulomb crystal could by applied to study various effects.
The sudden vacancy in a 2D crystal can create a topological defect, which has recently been discussed in the framework of the quasi-particle fracton model \cite{pretko2018fracton}.
Alternatively, one might probe structural defects known as 'kinks' \cite{mielenz2013trapping,pyka2013topological,brox2017spectroscopy}.
So far these have been investigated within the oscillating fields of an rf trap.
For future experiments, one might consider trapping a large 2D Coulomb crystal \cite{retzker2008double,shimshoni2011quantum} in a common state-dependent ODT.
Removing the center ion might create a defect while simultaneously eliminating contributions of micromotion.

The presented results can also be applied in the emergent field of ultracold atom-ion interactions \cite{harter2014cold,tomza2019cold}.
Hereby individual ions interact with an ensemble of atoms, allowing to investigate chemical reactions \cite{grier2009observation,hall2013light,saito2017characterization,joger2017observation}, 
spin dynamics \cite{ratschbacher2013decoherence,furst2018dynamics,sikorsky2018spin} and elastic collisions, aiming at the ultracold quantum regime \cite{feldker2020buffer,schmidt2020optical}.
All these measurements require constant reloading of a single, isolated ion.
The presented scheme enables a fast and reliable deterministic single ion source and might therefore be a usefull tool for any trapping platform.

\begin{center}
\textbf{Acknowledgements}
\end{center}

This project has received funding from the European Research Council (ERC) under the European Union’s Horizon 2020 research and innovation program (Grant No. 648330) and was supported by the Georg H. Endress foundation.
P.W., F.T. and T.S. acknowledge support from the DFG within the GRK 2079/1 program.
P.W. gratefully acknowledges financial support from the Studienstiftung des deutschen Volkes.
We are grateful for M. Debatin helping build the experimental setup.
We further thank J. Schmidt and D. Leibfried for fruitful discussions and T. Walker for constructive criticism of the manuscript.

\section{References}
\label{sec:ref}

\bibliography{deterministic}

\begin{thebibliography}{65}%
\makeatletter
\providecommand \@ifxundefined [1]{%
 \@ifx{#1\undefined}
}%
\providecommand \@ifnum [1]{%
 \ifnum #1\expandafter \@firstoftwo
 \else \expandafter \@secondoftwo
 \fi
}%
\providecommand \@ifx [1]{%
 \ifx #1\expandafter \@firstoftwo
 \else \expandafter \@secondoftwo
 \fi
}%
\providecommand \natexlab [1]{#1}%
\providecommand \enquote  [1]{``#1''}%
\providecommand \bibnamefont  [1]{#1}%
\providecommand \bibfnamefont [1]{#1}%
\providecommand \citenamefont [1]{#1}%
\providecommand \href@noop [0]{\@secondoftwo}%
\providecommand \href [0]{\begingroup \@sanitize@url \@href}%
\providecommand \@href[1]{\@@startlink{#1}\@@href}%
\providecommand \@@href[1]{\endgroup#1\@@endlink}%
\providecommand \@sanitize@url [0]{\catcode `\\12\catcode `\$12\catcode
  `\&12\catcode `\#12\catcode `\^12\catcode `\_12\catcode `\%12\relax}%
\providecommand \@@startlink[1]{}%
\providecommand \@@endlink[0]{}%
\providecommand \url  [0]{\begingroup\@sanitize@url \@url }%
\providecommand \@url [1]{\endgroup\@href {#1}{\urlprefix }}%
\providecommand \urlprefix  [0]{URL }%
\providecommand \Eprint [0]{\href }%
\providecommand \doibase [0]{http://dx.doi.org/}%
\providecommand \selectlanguage [0]{\@gobble}%
\providecommand \bibinfo  [0]{\@secondoftwo}%
\providecommand \bibfield  [0]{\@secondoftwo}%
\providecommand \translation [1]{[#1]}%
\providecommand \BibitemOpen [0]{}%
\providecommand \bibitemStop [0]{}%
\providecommand \bibitemNoStop [0]{.\EOS\space}%
\providecommand \EOS [0]{\spacefactor3000\relax}%
\providecommand \BibitemShut  [1]{\csname bibitem#1\endcsname}%
\let\auto@bib@innerbib\@empty
\bibitem [{\citenamefont {Paul}(1990)}]{paul1990electromagnetic}%
  \BibitemOpen
  \bibfield  {author} {\bibinfo {author} {\bibfnamefont {W.}~\bibnamefont
  {Paul}},\ }\href@noop {} {\bibfield  {journal} {\bibinfo  {journal}
  {Angewandte Chemie International Edition in English}\ }\textbf {\bibinfo
  {volume} {29}},\ \bibinfo {pages} {739} (\bibinfo {year} {1990})}\BibitemShut
  {NoStop}%
\bibitem [{\citenamefont {Dehmelt}(1991)}]{dehmelt1991experiments}%
  \BibitemOpen
  \bibfield  {author} {\bibinfo {author} {\bibfnamefont {H.}~\bibnamefont
  {Dehmelt}},\ }in\ \href@noop {} {\emph {\bibinfo {booktitle} {AIP Conference
  Proceedings}}},\ Vol.\ \bibinfo {volume} {233}\ (\bibinfo {organization}
  {American Institute of Physics},\ \bibinfo {year} {1991})\ pp.\ \bibinfo
  {pages} {28--45}\BibitemShut {NoStop}%
\bibitem [{\citenamefont {Wineland}(2013)}]{wineland2013nobel}%
  \BibitemOpen
  \bibfield  {author} {\bibinfo {author} {\bibfnamefont {D.~J.}\ \bibnamefont
  {Wineland}},\ }\href@noop {} {\bibfield  {journal} {\bibinfo  {journal}
  {Reviews of Modern Physics}\ }\textbf {\bibinfo {volume} {85}},\ \bibinfo
  {pages} {1103} (\bibinfo {year} {2013})}\BibitemShut {NoStop}%
\bibitem [{\citenamefont {Higgins}\ \emph {et~al.}(2017)\citenamefont
  {Higgins}, \citenamefont {Li}, \citenamefont {Pokorny}, \citenamefont
  {Zhang}, \citenamefont {Kress}, \citenamefont {Maier}, \citenamefont {Haag},
  \citenamefont {Bodart}, \citenamefont {Lesanovsky},\ and\ \citenamefont
  {Hennrich}}]{higgins2017single}%
  \BibitemOpen
  \bibfield  {author} {\bibinfo {author} {\bibfnamefont {G.}~\bibnamefont
  {Higgins}}, \bibinfo {author} {\bibfnamefont {W.}~\bibnamefont {Li}},
  \bibinfo {author} {\bibfnamefont {F.}~\bibnamefont {Pokorny}}, \bibinfo
  {author} {\bibfnamefont {C.}~\bibnamefont {Zhang}}, \bibinfo {author}
  {\bibfnamefont {F.}~\bibnamefont {Kress}}, \bibinfo {author} {\bibfnamefont
  {C.}~\bibnamefont {Maier}}, \bibinfo {author} {\bibfnamefont
  {J.}~\bibnamefont {Haag}}, \bibinfo {author} {\bibfnamefont {Q.}~\bibnamefont
  {Bodart}}, \bibinfo {author} {\bibfnamefont {I.}~\bibnamefont {Lesanovsky}},
  \ and\ \bibinfo {author} {\bibfnamefont {M.}~\bibnamefont {Hennrich}},\
  }\href@noop {} {\bibfield  {journal} {\bibinfo  {journal} {Physical Review
  X}\ }\textbf {\bibinfo {volume} {7}},\ \bibinfo {pages} {021038} (\bibinfo
  {year} {2017})}\BibitemShut {NoStop}%
\bibitem [{\citenamefont {Monroe}(2002)}]{monroe2002quantum}%
  \BibitemOpen
  \bibfield  {author} {\bibinfo {author} {\bibfnamefont {C.}~\bibnamefont
  {Monroe}},\ }\href@noop {} {\bibfield  {journal} {\bibinfo  {journal}
  {Nature}\ }\textbf {\bibinfo {volume} {416}},\ \bibinfo {pages} {238}
  (\bibinfo {year} {2002})}\BibitemShut {NoStop}%
\bibitem [{\citenamefont {Blatt}\ and\ \citenamefont
  {Roos}(2012)}]{blatt2012quantum}%
  \BibitemOpen
  \bibfield  {author} {\bibinfo {author} {\bibfnamefont {R.}~\bibnamefont
  {Blatt}}\ and\ \bibinfo {author} {\bibfnamefont {C.~F.}\ \bibnamefont
  {Roos}},\ }\href@noop {} {\bibfield  {journal} {\bibinfo  {journal} {Nature
  Physics}\ }\textbf {\bibinfo {volume} {8}},\ \bibinfo {pages} {277} (\bibinfo
  {year} {2012})}\BibitemShut {NoStop}%
\bibitem [{\citenamefont {Wineland}\ and\ \citenamefont
  {Leibfried}(2011)}]{wineland2011quantum}%
  \BibitemOpen
  \bibfield  {author} {\bibinfo {author} {\bibfnamefont {D.~J.}\ \bibnamefont
  {Wineland}}\ and\ \bibinfo {author} {\bibfnamefont {D.}~\bibnamefont
  {Leibfried}},\ }\href@noop {} {\bibfield  {journal} {\bibinfo  {journal}
  {Laser Physics Letters}\ }\textbf {\bibinfo {volume} {8}},\ \bibinfo {pages}
  {175} (\bibinfo {year} {2011})}\BibitemShut {NoStop}%
\bibitem [{\citenamefont {Ludlow}\ \emph {et~al.}(2015)\citenamefont {Ludlow},
  \citenamefont {Boyd}, \citenamefont {Ye}, \citenamefont {Peik},\ and\
  \citenamefont {Schmidt}}]{ludlow2015optical}%
  \BibitemOpen
  \bibfield  {author} {\bibinfo {author} {\bibfnamefont {A.~D.}\ \bibnamefont
  {Ludlow}}, \bibinfo {author} {\bibfnamefont {M.~M.}\ \bibnamefont {Boyd}},
  \bibinfo {author} {\bibfnamefont {J.}~\bibnamefont {Ye}}, \bibinfo {author}
  {\bibfnamefont {E.}~\bibnamefont {Peik}}, \ and\ \bibinfo {author}
  {\bibfnamefont {P.~O.}\ \bibnamefont {Schmidt}},\ }\href@noop {} {\bibfield
  {journal} {\bibinfo  {journal} {Reviews of Modern Physics}\ }\textbf
  {\bibinfo {volume} {87}},\ \bibinfo {pages} {637} (\bibinfo {year}
  {2015})}\BibitemShut {NoStop}%
\bibitem [{\citenamefont {Safronova}\ \emph {et~al.}(2014)\citenamefont
  {Safronova}, \citenamefont {Dzuba}, \citenamefont {Flambaum}, \citenamefont
  {Safronova}, \citenamefont {Porsev},\ and\ \citenamefont
  {Kozlov}}]{safronova2014highly}%
  \BibitemOpen
  \bibfield  {author} {\bibinfo {author} {\bibfnamefont {M.~S.}\ \bibnamefont
  {Safronova}}, \bibinfo {author} {\bibfnamefont {V.~A.}\ \bibnamefont
  {Dzuba}}, \bibinfo {author} {\bibfnamefont {V.~V.}\ \bibnamefont {Flambaum}},
  \bibinfo {author} {\bibfnamefont {U.~I.}\ \bibnamefont {Safronova}}, \bibinfo
  {author} {\bibfnamefont {S.~G.}\ \bibnamefont {Porsev}}, \ and\ \bibinfo
  {author} {\bibfnamefont {M.~G.}\ \bibnamefont {Kozlov}},\ }\href@noop {}
  {\bibfield  {journal} {\bibinfo  {journal} {Physical Review Letters}\
  }\textbf {\bibinfo {volume} {113}},\ \bibinfo {pages} {030801} (\bibinfo
  {year} {2014})}\BibitemShut {NoStop}%
\bibitem [{\citenamefont {Kozlov}\ \emph {et~al.}(2018)\citenamefont {Kozlov},
  \citenamefont {Safronova}, \citenamefont {Crespo Lopez-Urrutia},\ and\
  \citenamefont {Schmidt}}]{kozlov2018highly}%
  \BibitemOpen
  \bibfield  {author} {\bibinfo {author} {\bibfnamefont {M.~G.}\ \bibnamefont
  {Kozlov}}, \bibinfo {author} {\bibfnamefont {M.~S.}\ \bibnamefont
  {Safronova}}, \bibinfo {author} {\bibfnamefont {J.~R.}\ \bibnamefont {Crespo
  Lopez-Urrutia}}, \ and\ \bibinfo {author} {\bibfnamefont {P.~O.}\
  \bibnamefont {Schmidt}},\ }\href@noop {} {\bibfield  {journal} {\bibinfo
  {journal} {Reviews of Modern Physics}\ }\textbf {\bibinfo {volume} {90}},\
  \bibinfo {pages} {045005} (\bibinfo {year} {2018})}\BibitemShut {NoStop}%
\bibitem [{\citenamefont {Safronova}\ \emph {et~al.}(2013)\citenamefont
  {Safronova}, \citenamefont {Porsev}, \citenamefont {Safronova}, \citenamefont
  {Kozlov},\ and\ \citenamefont {Clark}}]{safronova2013blackbody}%
  \BibitemOpen
  \bibfield  {author} {\bibinfo {author} {\bibfnamefont {M.~S.}\ \bibnamefont
  {Safronova}}, \bibinfo {author} {\bibfnamefont {S.~G.}\ \bibnamefont
  {Porsev}}, \bibinfo {author} {\bibfnamefont {U.~I.}\ \bibnamefont
  {Safronova}}, \bibinfo {author} {\bibfnamefont {M.~G.}\ \bibnamefont
  {Kozlov}}, \ and\ \bibinfo {author} {\bibfnamefont {C.~W.}\ \bibnamefont
  {Clark}},\ }\href@noop {} {\bibfield  {journal} {\bibinfo  {journal}
  {Physical Review A}\ }\textbf {\bibinfo {volume} {87}},\ \bibinfo {pages}
  {012509} (\bibinfo {year} {2013})}\BibitemShut {NoStop}%
\bibitem [{\citenamefont {Beloy}\ \emph {et~al.}(2014)\citenamefont {Beloy},
  \citenamefont {Hinkley}, \citenamefont {Phillips}, \citenamefont {Sherman},
  \citenamefont {Schioppo}, \citenamefont {Lehman}, \citenamefont {Feldman},
  \citenamefont {Hanssen}, \citenamefont {Oates},\ and\ \citenamefont
  {Ludlow}}]{beloy2014atomic}%
  \BibitemOpen
  \bibfield  {author} {\bibinfo {author} {\bibfnamefont {K.}~\bibnamefont
  {Beloy}}, \bibinfo {author} {\bibfnamefont {N.}~\bibnamefont {Hinkley}},
  \bibinfo {author} {\bibfnamefont {N.~B.}\ \bibnamefont {Phillips}}, \bibinfo
  {author} {\bibfnamefont {J.~A.}\ \bibnamefont {Sherman}}, \bibinfo {author}
  {\bibfnamefont {M.}~\bibnamefont {Schioppo}}, \bibinfo {author}
  {\bibfnamefont {J.}~\bibnamefont {Lehman}}, \bibinfo {author} {\bibfnamefont
  {A.}~\bibnamefont {Feldman}}, \bibinfo {author} {\bibfnamefont {L.~M.}\
  \bibnamefont {Hanssen}}, \bibinfo {author} {\bibfnamefont {C.~W.}\
  \bibnamefont {Oates}}, \ and\ \bibinfo {author} {\bibfnamefont {A.~D.}\
  \bibnamefont {Ludlow}},\ }\href@noop {} {\bibfield  {journal} {\bibinfo
  {journal} {Physical Review Letters}\ }\textbf {\bibinfo {volume} {113}},\
  \bibinfo {pages} {260801} (\bibinfo {year} {2014})}\BibitemShut {NoStop}%
\bibitem [{\citenamefont {Huntemann}\ \emph {et~al.}(2016)\citenamefont
  {Huntemann}, \citenamefont {Sanner}, \citenamefont {Lipphardt}, \citenamefont
  {Tamm},\ and\ \citenamefont {Peik}}]{huntemann2016single}%
  \BibitemOpen
  \bibfield  {author} {\bibinfo {author} {\bibfnamefont {N.}~\bibnamefont
  {Huntemann}}, \bibinfo {author} {\bibfnamefont {C.}~\bibnamefont {Sanner}},
  \bibinfo {author} {\bibfnamefont {B.}~\bibnamefont {Lipphardt}}, \bibinfo
  {author} {\bibfnamefont {C.}~\bibnamefont {Tamm}}, \ and\ \bibinfo {author}
  {\bibfnamefont {E.}~\bibnamefont {Peik}},\ }\href@noop {} {\bibfield
  {journal} {\bibinfo  {journal} {Physical Review Letters}\ }\textbf {\bibinfo
  {volume} {116}},\ \bibinfo {pages} {063001} (\bibinfo {year}
  {2016})}\BibitemShut {NoStop}%
\bibitem [{\citenamefont {Beloy}\ \emph {et~al.}(2018)\citenamefont {Beloy},
  \citenamefont {Zhang}, \citenamefont {McGrew}, \citenamefont {Hinkley},
  \citenamefont {Yoon}, \citenamefont {Nicolodi}, \citenamefont {Fasano},
  \citenamefont {Schaffer}, \citenamefont {Brown},\ and\ \citenamefont
  {Ludlow}}]{beloy2018faraday}%
  \BibitemOpen
  \bibfield  {author} {\bibinfo {author} {\bibfnamefont {K.}~\bibnamefont
  {Beloy}}, \bibinfo {author} {\bibfnamefont {X.}~\bibnamefont {Zhang}},
  \bibinfo {author} {\bibfnamefont {W.~F.}\ \bibnamefont {McGrew}}, \bibinfo
  {author} {\bibfnamefont {N.}~\bibnamefont {Hinkley}}, \bibinfo {author}
  {\bibfnamefont {T.~H.}\ \bibnamefont {Yoon}}, \bibinfo {author}
  {\bibfnamefont {D.}~\bibnamefont {Nicolodi}}, \bibinfo {author}
  {\bibfnamefont {R.~J.}\ \bibnamefont {Fasano}}, \bibinfo {author}
  {\bibfnamefont {S.~A.}\ \bibnamefont {Schaffer}}, \bibinfo {author}
  {\bibfnamefont {R.~C.}\ \bibnamefont {Brown}}, \ and\ \bibinfo {author}
  {\bibfnamefont {A.~D.}\ \bibnamefont {Ludlow}},\ }\href@noop {} {\bibfield
  {journal} {\bibinfo  {journal} {Physical Review Letters}\ }\textbf {\bibinfo
  {volume} {120}},\ \bibinfo {pages} {183201} (\bibinfo {year}
  {2018})}\BibitemShut {NoStop}%
\bibitem [{\citenamefont {Guggemos}\ \emph
  {et~al.}(2015{\natexlab{a}})\citenamefont {Guggemos}, \citenamefont
  {Heinrich}, \citenamefont {Herrera-Sancho}, \citenamefont {Blatt},\ and\
  \citenamefont {Roos}}]{GuggemosHeinrichHerrera-SanchoEtAl2015}%
  \BibitemOpen
  \bibfield  {author} {\bibinfo {author} {\bibfnamefont {M.}~\bibnamefont
  {Guggemos}}, \bibinfo {author} {\bibfnamefont {D.}~\bibnamefont {Heinrich}},
  \bibinfo {author} {\bibfnamefont {O.}~\bibnamefont {Herrera-Sancho}},
  \bibinfo {author} {\bibfnamefont {R.}~\bibnamefont {Blatt}}, \ and\ \bibinfo
  {author} {\bibfnamefont {C.}~\bibnamefont {Roos}},\ }\href@noop {} {\bibfield
   {journal} {\bibinfo  {journal} {New Journal of Physics}\ }\textbf {\bibinfo
  {volume} {17}},\ \bibinfo {pages} {103001} (\bibinfo {year}
  {2015}{\natexlab{a}})}\BibitemShut {NoStop}%
\bibitem [{\citenamefont {Drewsen}\ \emph {et~al.}(2004)\citenamefont
  {Drewsen}, \citenamefont {Mortensen}, \citenamefont {Martinussen},
  \citenamefont {Staanum},\ and\ \citenamefont
  {S{\o}rensen}}]{drewsen2004nondestructive}%
  \BibitemOpen
  \bibfield  {author} {\bibinfo {author} {\bibfnamefont {M.}~\bibnamefont
  {Drewsen}}, \bibinfo {author} {\bibfnamefont {A.}~\bibnamefont {Mortensen}},
  \bibinfo {author} {\bibfnamefont {R.}~\bibnamefont {Martinussen}}, \bibinfo
  {author} {\bibfnamefont {P.}~\bibnamefont {Staanum}}, \ and\ \bibinfo
  {author} {\bibfnamefont {J.~L.}\ \bibnamefont {S{\o}rensen}},\ }\href@noop {}
  {\bibfield  {journal} {\bibinfo  {journal} {Physical Review Letters}\
  }\textbf {\bibinfo {volume} {93}},\ \bibinfo {pages} {243201} (\bibinfo
  {year} {2004})}\BibitemShut {NoStop}%
\bibitem [{\citenamefont {N{\"a}gerl}\ \emph {et~al.}(1998)\citenamefont
  {N{\"a}gerl}, \citenamefont {Leibfried}, \citenamefont {Schmidt-Kaler},
  \citenamefont {Eschner},\ and\ \citenamefont {Blatt}}]{nagerl1998coherent}%
  \BibitemOpen
  \bibfield  {author} {\bibinfo {author} {\bibfnamefont {H.}~\bibnamefont
  {N{\"a}gerl}}, \bibinfo {author} {\bibfnamefont {D.}~\bibnamefont
  {Leibfried}}, \bibinfo {author} {\bibfnamefont {F.}~\bibnamefont
  {Schmidt-Kaler}}, \bibinfo {author} {\bibfnamefont {J.}~\bibnamefont
  {Eschner}}, \ and\ \bibinfo {author} {\bibfnamefont {R.}~\bibnamefont
  {Blatt}},\ }\href@noop {} {\bibfield  {journal} {\bibinfo  {journal} {Optics
  Express}\ }\textbf {\bibinfo {volume} {3}},\ \bibinfo {pages} {89} (\bibinfo
  {year} {1998})}\BibitemShut {NoStop}%
\bibitem [{\citenamefont {Razvi}\ \emph {et~al.}(1998)\citenamefont {Razvi},
  \citenamefont {Chu}, \citenamefont {Alheit}, \citenamefont {Werth},\ and\
  \citenamefont {Blumel}}]{razvi1998fractional}%
  \BibitemOpen
  \bibfield  {author} {\bibinfo {author} {\bibfnamefont {M.~A.~N.}\
  \bibnamefont {Razvi}}, \bibinfo {author} {\bibfnamefont {X.~Z.}\ \bibnamefont
  {Chu}}, \bibinfo {author} {\bibfnamefont {R.}~\bibnamefont {Alheit}},
  \bibinfo {author} {\bibfnamefont {G.}~\bibnamefont {Werth}}, \ and\ \bibinfo
  {author} {\bibfnamefont {R.}~\bibnamefont {Blumel}},\ }\href@noop {}
  {\bibfield  {journal} {\bibinfo  {journal} {Physical Review A}\ }\textbf
  {\bibinfo {volume} {58}},\ \bibinfo {pages} {R34} (\bibinfo {year}
  {1998})}\BibitemShut {NoStop}%
\bibitem [{\citenamefont {Sudakov}\ \emph {et~al.}(2000)\citenamefont
  {Sudakov}, \citenamefont {Konenkov}, \citenamefont {Douglas},\ and\
  \citenamefont {Glebova}}]{sudakov2000excitation}%
  \BibitemOpen
  \bibfield  {author} {\bibinfo {author} {\bibfnamefont {M.}~\bibnamefont
  {Sudakov}}, \bibinfo {author} {\bibfnamefont {N.}~\bibnamefont {Konenkov}},
  \bibinfo {author} {\bibfnamefont {D.}~\bibnamefont {Douglas}}, \ and\
  \bibinfo {author} {\bibfnamefont {T.}~\bibnamefont {Glebova}},\ }\href@noop
  {} {\bibfield  {journal} {\bibinfo  {journal} {Journal of the American
  Society for Mass Spectrometry}\ }\textbf {\bibinfo {volume} {11}},\ \bibinfo
  {pages} {10} (\bibinfo {year} {2000})}\BibitemShut {NoStop}%
\bibitem [{\citenamefont {Schmidt}\ \emph
  {et~al.}(2020{\natexlab{a}})\citenamefont {Schmidt}, \citenamefont
  {H{\"o}nig}, \citenamefont {Weckesser}, \citenamefont {Thielemann},
  \citenamefont {Schaetz},\ and\ \citenamefont {Karpa}}]{schmidt2020mass}%
  \BibitemOpen
  \bibfield  {author} {\bibinfo {author} {\bibfnamefont {J.}~\bibnamefont
  {Schmidt}}, \bibinfo {author} {\bibfnamefont {D.}~\bibnamefont {H{\"o}nig}},
  \bibinfo {author} {\bibfnamefont {P.}~\bibnamefont {Weckesser}}, \bibinfo
  {author} {\bibfnamefont {F.}~\bibnamefont {Thielemann}}, \bibinfo {author}
  {\bibfnamefont {T.}~\bibnamefont {Schaetz}}, \ and\ \bibinfo {author}
  {\bibfnamefont {L.}~\bibnamefont {Karpa}},\ }\href@noop {} {\bibfield
  {journal} {\bibinfo  {journal} {Applied Physics B}\ }\textbf {\bibinfo
  {volume} {126}},\ \bibinfo {pages} {1} (\bibinfo {year}
  {2020}{\natexlab{a}})}\BibitemShut {NoStop}%
\bibitem [{\citenamefont {Schneider}\ \emph {et~al.}(2010)\citenamefont
  {Schneider}, \citenamefont {Enderlein}, \citenamefont {Huber},\ and\
  \citenamefont {Sch{\"a}tz}}]{schneider2010optical}%
  \BibitemOpen
  \bibfield  {author} {\bibinfo {author} {\bibfnamefont {C.}~\bibnamefont
  {Schneider}}, \bibinfo {author} {\bibfnamefont {M.}~\bibnamefont
  {Enderlein}}, \bibinfo {author} {\bibfnamefont {T.}~\bibnamefont {Huber}}, \
  and\ \bibinfo {author} {\bibfnamefont {T.}~\bibnamefont {Sch{\"a}tz}},\
  }\href@noop {} {\bibfield  {journal} {\bibinfo  {journal} {Nature Photonics}\
  }\textbf {\bibinfo {volume} {4}},\ \bibinfo {pages} {772} (\bibinfo {year}
  {2010})}\BibitemShut {NoStop}%
\bibitem [{\citenamefont {Huber}\ \emph {et~al.}(2014)\citenamefont {Huber},
  \citenamefont {Lambrecht}, \citenamefont {Schmidt}, \citenamefont {Karpa},\
  and\ \citenamefont {Schaetz}}]{huber2014far}%
  \BibitemOpen
  \bibfield  {author} {\bibinfo {author} {\bibfnamefont {T.}~\bibnamefont
  {Huber}}, \bibinfo {author} {\bibfnamefont {A.}~\bibnamefont {Lambrecht}},
  \bibinfo {author} {\bibfnamefont {J.}~\bibnamefont {Schmidt}}, \bibinfo
  {author} {\bibfnamefont {L.}~\bibnamefont {Karpa}}, \ and\ \bibinfo {author}
  {\bibfnamefont {T.}~\bibnamefont {Schaetz}},\ }\href@noop {} {\bibfield
  {journal} {\bibinfo  {journal} {Nature Communications}\ }\textbf {\bibinfo
  {volume} {5}},\ \bibinfo {pages} {5587} (\bibinfo {year} {2014})}\BibitemShut
  {NoStop}%
\bibitem [{\citenamefont {Lambrecht}\ \emph {et~al.}(2017)\citenamefont
  {Lambrecht}, \citenamefont {Schmidt}, \citenamefont {Weckesser},
  \citenamefont {Debatin}, \citenamefont {Karpa},\ and\ \citenamefont
  {Schaetz}}]{lambrecht2017long}%
  \BibitemOpen
  \bibfield  {author} {\bibinfo {author} {\bibfnamefont {A.}~\bibnamefont
  {Lambrecht}}, \bibinfo {author} {\bibfnamefont {J.}~\bibnamefont {Schmidt}},
  \bibinfo {author} {\bibfnamefont {P.}~\bibnamefont {Weckesser}}, \bibinfo
  {author} {\bibfnamefont {M.}~\bibnamefont {Debatin}}, \bibinfo {author}
  {\bibfnamefont {L.}~\bibnamefont {Karpa}}, \ and\ \bibinfo {author}
  {\bibfnamefont {T.}~\bibnamefont {Schaetz}},\ }\href@noop {} {\bibfield
  {journal} {\bibinfo  {journal} {Nature Photonics}\ }\textbf {\bibinfo
  {volume} {11}},\ \bibinfo {pages} {704} (\bibinfo {year} {2017})}\BibitemShut
  {NoStop}%
\bibitem [{\citenamefont {Schaetz}(2017)}]{schaetz2017trapping}%
  \BibitemOpen
  \bibfield  {author} {\bibinfo {author} {\bibfnamefont {T.}~\bibnamefont
  {Schaetz}},\ }\href@noop {} {\bibfield  {journal} {\bibinfo  {journal}
  {Journal of Physics B: Atomic, Molecular and Optical Physics}\ }\textbf
  {\bibinfo {volume} {50}},\ \bibinfo {pages} {102001} (\bibinfo {year}
  {2017})}\BibitemShut {NoStop}%
\bibitem [{\citenamefont {Schmidt}\ \emph {et~al.}(2018)\citenamefont
  {Schmidt}, \citenamefont {Lambrecht}, \citenamefont {Weckesser},
  \citenamefont {Debatin}, \citenamefont {Karpa},\ and\ \citenamefont
  {Schaetz}}]{schmidt2018optical}%
  \BibitemOpen
  \bibfield  {author} {\bibinfo {author} {\bibfnamefont {J.}~\bibnamefont
  {Schmidt}}, \bibinfo {author} {\bibfnamefont {A.}~\bibnamefont {Lambrecht}},
  \bibinfo {author} {\bibfnamefont {P.}~\bibnamefont {Weckesser}}, \bibinfo
  {author} {\bibfnamefont {M.}~\bibnamefont {Debatin}}, \bibinfo {author}
  {\bibfnamefont {L.}~\bibnamefont {Karpa}}, \ and\ \bibinfo {author}
  {\bibfnamefont {T.}~\bibnamefont {Schaetz}},\ }\href@noop {} {\bibfield
  {journal} {\bibinfo  {journal} {Physical Review X}\ }\textbf {\bibinfo
  {volume} {8}},\ \bibinfo {pages} {021028} (\bibinfo {year}
  {2018})}\BibitemShut {NoStop}%
\bibitem [{\citenamefont {Karpa}(2019)}]{karpa2019trapping}%
  \BibitemOpen
  \bibfield  {author} {\bibinfo {author} {\bibfnamefont {L.}~\bibnamefont
  {Karpa}},\ }\href@noop {} {\emph {\bibinfo {title} {Trapping single ions and
  Coulomb crystals with light fields}}}\ (\bibinfo  {publisher} {Springer},\
  \bibinfo {year} {2019})\BibitemShut {NoStop}%
\bibitem [{\citenamefont {Cormick}\ \emph {et~al.}(2011)\citenamefont
  {Cormick}, \citenamefont {Schaetz},\ and\ \citenamefont
  {Morigi}}]{cormick2011trapping}%
  \BibitemOpen
  \bibfield  {author} {\bibinfo {author} {\bibfnamefont {C.}~\bibnamefont
  {Cormick}}, \bibinfo {author} {\bibfnamefont {T.}~\bibnamefont {Schaetz}}, \
  and\ \bibinfo {author} {\bibfnamefont {G.}~\bibnamefont {Morigi}},\
  }\href@noop {} {\bibfield  {journal} {\bibinfo  {journal} {New Journal of
  Physics}\ }\textbf {\bibinfo {volume} {13}},\ \bibinfo {pages} {043019}
  (\bibinfo {year} {2011})}\BibitemShut {NoStop}%
\bibitem [{\citenamefont {Grimm}\ \emph {et~al.}(2000)\citenamefont {Grimm},
  \citenamefont {Weidem{\"u}ller},\ and\ \citenamefont
  {Ovchinnikov}}]{grimm2000optical}%
  \BibitemOpen
  \bibfield  {author} {\bibinfo {author} {\bibfnamefont {R.}~\bibnamefont
  {Grimm}}, \bibinfo {author} {\bibfnamefont {M.}~\bibnamefont
  {Weidem{\"u}ller}}, \ and\ \bibinfo {author} {\bibfnamefont {Y.~B.}\
  \bibnamefont {Ovchinnikov}},\ }in\ \href@noop {} {\emph {\bibinfo {booktitle}
  {Advances in atomic, molecular, and optical physics}}},\ Vol.~\bibinfo
  {volume} {42}\ (\bibinfo  {publisher} {Elsevier},\ \bibinfo {year} {2000})\
  pp.\ \bibinfo {pages} {95--170}\BibitemShut {NoStop}%
\bibitem [{\citenamefont {Blatt}\ and\ \citenamefont
  {Wineland}(2008)}]{blatt2008entangled}%
  \BibitemOpen
  \bibfield  {author} {\bibinfo {author} {\bibfnamefont {R.}~\bibnamefont
  {Blatt}}\ and\ \bibinfo {author} {\bibfnamefont {D.}~\bibnamefont
  {Wineland}},\ }\href@noop {} {\bibfield  {journal} {\bibinfo  {journal}
  {Nature}\ }\textbf {\bibinfo {volume} {453}},\ \bibinfo {pages} {1008}
  (\bibinfo {year} {2008})}\BibitemShut {NoStop}%
\bibitem [{\citenamefont {Leibfried}\ \emph {et~al.}(2003)\citenamefont
  {Leibfried}, \citenamefont {DeMarco}, \citenamefont {Meyer}, \citenamefont
  {Lucas}, \citenamefont {Barrett}, \citenamefont {Britton}, \citenamefont
  {Itano}, \citenamefont {Jelenkovi{\'c}}, \citenamefont {Langer},
  \citenamefont {Rosenband} \emph {et~al.}}]{leibfried2003experimental}%
  \BibitemOpen
  \bibfield  {author} {\bibinfo {author} {\bibfnamefont {D.}~\bibnamefont
  {Leibfried}}, \bibinfo {author} {\bibfnamefont {B.}~\bibnamefont {DeMarco}},
  \bibinfo {author} {\bibfnamefont {V.}~\bibnamefont {Meyer}}, \bibinfo
  {author} {\bibfnamefont {D.}~\bibnamefont {Lucas}}, \bibinfo {author}
  {\bibfnamefont {M.}~\bibnamefont {Barrett}}, \bibinfo {author} {\bibfnamefont
  {J.}~\bibnamefont {Britton}}, \bibinfo {author} {\bibfnamefont {W.~M.}\
  \bibnamefont {Itano}}, \bibinfo {author} {\bibfnamefont {B.}~\bibnamefont
  {Jelenkovi{\'c}}}, \bibinfo {author} {\bibfnamefont {C.}~\bibnamefont
  {Langer}}, \bibinfo {author} {\bibfnamefont {T.}~\bibnamefont {Rosenband}},
  \emph {et~al.},\ }\href@noop {} {\bibfield  {journal} {\bibinfo  {journal}
  {Nature}\ }\textbf {\bibinfo {volume} {422}},\ \bibinfo {pages} {412}
  (\bibinfo {year} {2003})}\BibitemShut {NoStop}%
\bibitem [{\citenamefont {Porras}\ and\ \citenamefont
  {Cirac}(2004)}]{porras2004effective}%
  \BibitemOpen
  \bibfield  {author} {\bibinfo {author} {\bibfnamefont {D.}~\bibnamefont
  {Porras}}\ and\ \bibinfo {author} {\bibfnamefont {J.~I.}\ \bibnamefont
  {Cirac}},\ }\href@noop {} {\bibfield  {journal} {\bibinfo  {journal}
  {Physical Review Letters}\ }\textbf {\bibinfo {volume} {92}},\ \bibinfo
  {pages} {207901} (\bibinfo {year} {2004})}\BibitemShut {NoStop}%
\bibitem [{\citenamefont {Friedenauer}\ \emph {et~al.}(2008)\citenamefont
  {Friedenauer}, \citenamefont {Schmitz}, \citenamefont {Glueckert},
  \citenamefont {Porras},\ and\ \citenamefont
  {Sch{\"a}tz}}]{friedenauer2008simulating}%
  \BibitemOpen
  \bibfield  {author} {\bibinfo {author} {\bibfnamefont {A.}~\bibnamefont
  {Friedenauer}}, \bibinfo {author} {\bibfnamefont {H.}~\bibnamefont
  {Schmitz}}, \bibinfo {author} {\bibfnamefont {J.~T.}\ \bibnamefont
  {Glueckert}}, \bibinfo {author} {\bibfnamefont {D.}~\bibnamefont {Porras}}, \
  and\ \bibinfo {author} {\bibfnamefont {T.}~\bibnamefont {Sch{\"a}tz}},\
  }\href@noop {} {\bibfield  {journal} {\bibinfo  {journal} {Nature Physics}\
  }\textbf {\bibinfo {volume} {4}},\ \bibinfo {pages} {757} (\bibinfo {year}
  {2008})}\BibitemShut {NoStop}%
\bibitem [{\citenamefont {Linnet}\ \emph {et~al.}(2012)\citenamefont {Linnet},
  \citenamefont {Leroux}, \citenamefont {Marciante}, \citenamefont {Dantan},\
  and\ \citenamefont {Drewsen}}]{linnet2012pinning}%
  \BibitemOpen
  \bibfield  {author} {\bibinfo {author} {\bibfnamefont {R.~B.}\ \bibnamefont
  {Linnet}}, \bibinfo {author} {\bibfnamefont {I.~D.}\ \bibnamefont {Leroux}},
  \bibinfo {author} {\bibfnamefont {M.}~\bibnamefont {Marciante}}, \bibinfo
  {author} {\bibfnamefont {A.}~\bibnamefont {Dantan}}, \ and\ \bibinfo {author}
  {\bibfnamefont {M.}~\bibnamefont {Drewsen}},\ }\href@noop {} {\bibfield
  {journal} {\bibinfo  {journal} {Physical Review Letters}\ }\textbf {\bibinfo
  {volume} {109}},\ \bibinfo {pages} {233005} (\bibinfo {year}
  {2012})}\BibitemShut {NoStop}%
\bibitem [{\citenamefont {Karpa}\ \emph {et~al.}(2013)\citenamefont {Karpa},
  \citenamefont {Bylinskii}, \citenamefont {Gangloff}, \citenamefont {Cetina},\
  and\ \citenamefont {Vuleti{\'c}}}]{karpa2013suppression}%
  \BibitemOpen
  \bibfield  {author} {\bibinfo {author} {\bibfnamefont {L.}~\bibnamefont
  {Karpa}}, \bibinfo {author} {\bibfnamefont {A.}~\bibnamefont {Bylinskii}},
  \bibinfo {author} {\bibfnamefont {D.}~\bibnamefont {Gangloff}}, \bibinfo
  {author} {\bibfnamefont {M.}~\bibnamefont {Cetina}}, \ and\ \bibinfo {author}
  {\bibfnamefont {V.}~\bibnamefont {Vuleti{\'c}}},\ }\href@noop {} {\bibfield
  {journal} {\bibinfo  {journal} {Physical Review Letters}\ }\textbf {\bibinfo
  {volume} {111}},\ \bibinfo {pages} {163002} (\bibinfo {year}
  {2013})}\BibitemShut {NoStop}%
\bibitem [{\citenamefont {Laupretre}\ \emph {et~al.}(2019)\citenamefont
  {Laupretre}, \citenamefont {Linnet}, \citenamefont {Leroux}, \citenamefont
  {Landa}, \citenamefont {Dantan},\ and\ \citenamefont
  {Drewsen}}]{laupretre2019controlling}%
  \BibitemOpen
  \bibfield  {author} {\bibinfo {author} {\bibfnamefont {T.}~\bibnamefont
  {Laupretre}}, \bibinfo {author} {\bibfnamefont {R.~B.}\ \bibnamefont
  {Linnet}}, \bibinfo {author} {\bibfnamefont {I.~D.}\ \bibnamefont {Leroux}},
  \bibinfo {author} {\bibfnamefont {H.}~\bibnamefont {Landa}}, \bibinfo
  {author} {\bibfnamefont {A.}~\bibnamefont {Dantan}}, \ and\ \bibinfo {author}
  {\bibfnamefont {M.}~\bibnamefont {Drewsen}},\ }\href@noop {} {\bibfield
  {journal} {\bibinfo  {journal} {Physical Review A}\ }\textbf {\bibinfo
  {volume} {99}},\ \bibinfo {pages} {031401(R)} (\bibinfo {year}
  {2019})}\BibitemShut {NoStop}%
\bibitem [{\citenamefont {Bylinskii}\ \emph {et~al.}(2016)\citenamefont
  {Bylinskii}, \citenamefont {Gangloff}, \citenamefont {Counts},\ and\
  \citenamefont {Vuleti{\'c}}}]{bylinskii2016observation}%
  \BibitemOpen
  \bibfield  {author} {\bibinfo {author} {\bibfnamefont {A.}~\bibnamefont
  {Bylinskii}}, \bibinfo {author} {\bibfnamefont {D.}~\bibnamefont {Gangloff}},
  \bibinfo {author} {\bibfnamefont {I.}~\bibnamefont {Counts}}, \ and\ \bibinfo
  {author} {\bibfnamefont {V.}~\bibnamefont {Vuleti{\'c}}},\ }\href@noop {}
  {\bibfield  {journal} {\bibinfo  {journal} {Nature materials}\ }\textbf
  {\bibinfo {volume} {15}},\ \bibinfo {pages} {717} (\bibinfo {year}
  {2016})}\BibitemShut {NoStop}%
\bibitem [{\citenamefont {Leschhorn}\ \emph {et~al.}(2012)\citenamefont
  {Leschhorn}, \citenamefont {Hasegawa},\ and\ \citenamefont
  {Schaetz}}]{leschhorn2012efficient}%
  \BibitemOpen
  \bibfield  {author} {\bibinfo {author} {\bibfnamefont {G.}~\bibnamefont
  {Leschhorn}}, \bibinfo {author} {\bibfnamefont {T.}~\bibnamefont {Hasegawa}},
  \ and\ \bibinfo {author} {\bibfnamefont {T.}~\bibnamefont {Schaetz}},\
  }\href@noop {} {\bibfield  {journal} {\bibinfo  {journal} {Applied Physics
  B}\ }\textbf {\bibinfo {volume} {108}},\ \bibinfo {pages} {159} (\bibinfo
  {year} {2012})}\BibitemShut {NoStop}%
\bibitem [{\citenamefont {N{\"a}gerl}\ \emph {et~al.}(1999)\citenamefont
  {N{\"a}gerl}, \citenamefont {Leibfried}, \citenamefont {Rohde}, \citenamefont
  {Thalhammer}, \citenamefont {Eschner}, \citenamefont {Schmidt-Kaler},\ and\
  \citenamefont {Blatt}}]{naegerl1999laser}%
  \BibitemOpen
  \bibfield  {author} {\bibinfo {author} {\bibfnamefont {H.~C.}\ \bibnamefont
  {N{\"a}gerl}}, \bibinfo {author} {\bibfnamefont {D.}~\bibnamefont
  {Leibfried}}, \bibinfo {author} {\bibfnamefont {H.}~\bibnamefont {Rohde}},
  \bibinfo {author} {\bibfnamefont {G.}~\bibnamefont {Thalhammer}}, \bibinfo
  {author} {\bibfnamefont {J.}~\bibnamefont {Eschner}}, \bibinfo {author}
  {\bibfnamefont {F.}~\bibnamefont {Schmidt-Kaler}}, \ and\ \bibinfo {author}
  {\bibfnamefont {R.}~\bibnamefont {Blatt}},\ }\href@noop {} {\bibfield
  {journal} {\bibinfo  {journal} {Physical Review A}\ }\textbf {\bibinfo
  {volume} {60}},\ \bibinfo {pages} {145} (\bibinfo {year} {1999})}\BibitemShut
  {NoStop}%
\bibitem [{\citenamefont {Auchter}\ \emph {et~al.}(2014)\citenamefont
  {Auchter}, \citenamefont {Noel}, \citenamefont {Hoffman}, \citenamefont
  {Williams},\ and\ \citenamefont {Blinov}}]{auchter2014measurement}%
  \BibitemOpen
  \bibfield  {author} {\bibinfo {author} {\bibfnamefont {C.}~\bibnamefont
  {Auchter}}, \bibinfo {author} {\bibfnamefont {T.~W.}\ \bibnamefont {Noel}},
  \bibinfo {author} {\bibfnamefont {M.~R.}\ \bibnamefont {Hoffman}}, \bibinfo
  {author} {\bibfnamefont {S.~R.}\ \bibnamefont {Williams}}, \ and\ \bibinfo
  {author} {\bibfnamefont {B.~B.}\ \bibnamefont {Blinov}},\ }\href@noop {}
  {\bibfield  {journal} {\bibinfo  {journal} {Physical Review A}\ }\textbf
  {\bibinfo {volume} {90}},\ \bibinfo {pages} {060501(R)} (\bibinfo {year}
  {2014})}\BibitemShut {NoStop}%
\bibitem [{\citenamefont {Maxwell}(1873)}]{maxwell1873treatise}%
  \BibitemOpen
  \bibfield  {author} {\bibinfo {author} {\bibfnamefont {J.~C.}\ \bibnamefont
  {Maxwell}},\ }\href@noop {} {\emph {\bibinfo {title} {A treatise on
  electricity and magnetism}}},\ Vol.~\bibinfo {volume} {1}\ (\bibinfo
  {publisher} {Clarendon press},\ \bibinfo {year} {1873})\BibitemShut {NoStop}%
\bibitem [{\citenamefont {James}(1998)}]{james1998quantum}%
  \BibitemOpen
  \bibfield  {author} {\bibinfo {author} {\bibfnamefont {D.~F.}\ \bibnamefont
  {James}},\ }\href@noop {} {\bibfield  {journal} {\bibinfo  {journal} {Applied
  Physics B: Lasers and Optics}\ }\textbf {\bibinfo {volume} {66}},\ \bibinfo
  {pages} {181} (\bibinfo {year} {1998})}\BibitemShut {NoStop}%
\bibitem [{\citenamefont {Schneider}\ \emph {et~al.}(2012)\citenamefont
  {Schneider}, \citenamefont {Enderlein}, \citenamefont {Huber}, \citenamefont
  {D\"urr},\ and\ \citenamefont {Schaetz}}]{Schneider2012influence}%
  \BibitemOpen
  \bibfield  {author} {\bibinfo {author} {\bibfnamefont {C.}~\bibnamefont
  {Schneider}}, \bibinfo {author} {\bibfnamefont {M.}~\bibnamefont
  {Enderlein}}, \bibinfo {author} {\bibfnamefont {T.}~\bibnamefont {Huber}},
  \bibinfo {author} {\bibfnamefont {S.}~\bibnamefont {D\"urr}}, \ and\ \bibinfo
  {author} {\bibfnamefont {T.}~\bibnamefont {Schaetz}},\ }\href {\doibase
  10.1103/PhysRevA.85.013422} {\bibfield  {journal} {\bibinfo  {journal} {Phys.
  Rev. A}\ }\textbf {\bibinfo {volume} {85}},\ \bibinfo {pages} {013422}
  (\bibinfo {year} {2012})}\BibitemShut {NoStop}%
\bibitem [{\citenamefont {Kaur}\ \emph {et~al.}(2015)\citenamefont {Kaur},
  \citenamefont {Singh}, \citenamefont {Arora},\ and\ \citenamefont
  {Sahoo}}]{kaur2015magic}%
  \BibitemOpen
  \bibfield  {author} {\bibinfo {author} {\bibfnamefont {J.}~\bibnamefont
  {Kaur}}, \bibinfo {author} {\bibfnamefont {S.}~\bibnamefont {Singh}},
  \bibinfo {author} {\bibfnamefont {B.}~\bibnamefont {Arora}}, \ and\ \bibinfo
  {author} {\bibfnamefont {B.~K.}\ \bibnamefont {Sahoo}},\ }\href@noop {}
  {\bibfield  {journal} {\bibinfo  {journal} {Physical Review A}\ }\textbf
  {\bibinfo {volume} {92}},\ \bibinfo {pages} {031402(R)} (\bibinfo {year}
  {2015})}\BibitemShut {NoStop}%
\bibitem [{\citenamefont {De~Munshi}\ \emph {et~al.}(2015)\citenamefont
  {De~Munshi}, \citenamefont {Dutta}, \citenamefont {Rebhi},\ and\
  \citenamefont {Mukherjee}}]{de2015precision}%
  \BibitemOpen
  \bibfield  {author} {\bibinfo {author} {\bibfnamefont {D.}~\bibnamefont
  {De~Munshi}}, \bibinfo {author} {\bibfnamefont {T.}~\bibnamefont {Dutta}},
  \bibinfo {author} {\bibfnamefont {R.}~\bibnamefont {Rebhi}}, \ and\ \bibinfo
  {author} {\bibfnamefont {M.}~\bibnamefont {Mukherjee}},\ }\href@noop {}
  {\bibfield  {journal} {\bibinfo  {journal} {Physical Review A}\ }\textbf
  {\bibinfo {volume} {91}},\ \bibinfo {pages} {040501(R)} (\bibinfo {year}
  {2015})}\BibitemShut {NoStop}%
\bibitem [{\citenamefont {Arnold}\ \emph {et~al.}(2019)\citenamefont {Arnold},
  \citenamefont {Chanu}, \citenamefont {Kaewuam}, \citenamefont {Tan},
  \citenamefont {Yeo}, \citenamefont {Zhang}, \citenamefont {Safronova},\ and\
  \citenamefont {Barrett}}]{arnold2019measurements}%
  \BibitemOpen
  \bibfield  {author} {\bibinfo {author} {\bibfnamefont {K.~J.}\ \bibnamefont
  {Arnold}}, \bibinfo {author} {\bibfnamefont {S.~R.}\ \bibnamefont {Chanu}},
  \bibinfo {author} {\bibfnamefont {R.}~\bibnamefont {Kaewuam}}, \bibinfo
  {author} {\bibfnamefont {T.~R.}\ \bibnamefont {Tan}}, \bibinfo {author}
  {\bibfnamefont {L.}~\bibnamefont {Yeo}}, \bibinfo {author} {\bibfnamefont
  {Z.}~\bibnamefont {Zhang}}, \bibinfo {author} {\bibfnamefont {M.~S.}\
  \bibnamefont {Safronova}}, \ and\ \bibinfo {author} {\bibfnamefont {M.~D.}\
  \bibnamefont {Barrett}},\ }\href@noop {} {\bibfield  {journal} {\bibinfo
  {journal} {Physical Review A}\ }\textbf {\bibinfo {volume} {100}},\ \bibinfo
  {pages} {032503} (\bibinfo {year} {2019})}\BibitemShut {NoStop}%
\bibitem [{\citenamefont {Guggemos}\ \emph
  {et~al.}(2015{\natexlab{b}})\citenamefont {Guggemos}, \citenamefont
  {Heinrich}, \citenamefont {Herrera-Sancho}, \citenamefont {Blatt},\ and\
  \citenamefont {Roos}}]{GuggemosHeinrichHerrera-SanchoEtAl2015a}%
  \BibitemOpen
  \bibfield  {author} {\bibinfo {author} {\bibfnamefont {M.}~\bibnamefont
  {Guggemos}}, \bibinfo {author} {\bibfnamefont {D.}~\bibnamefont {Heinrich}},
  \bibinfo {author} {\bibfnamefont {O.}~\bibnamefont {Herrera-Sancho}},
  \bibinfo {author} {\bibfnamefont {R.}~\bibnamefont {Blatt}}, \ and\ \bibinfo
  {author} {\bibfnamefont {C.}~\bibnamefont {Roos}},\ }\href@noop {} {\bibfield
   {journal} {\bibinfo  {journal} {New Journal of Physics}\ }\textbf {\bibinfo
  {volume} {17}},\ \bibinfo {pages} {103001} (\bibinfo {year}
  {2015}{\natexlab{b}})}\BibitemShut {NoStop}%
\bibitem [{\citenamefont {Hendricks}\ \emph {et~al.}(2007)\citenamefont
  {Hendricks}, \citenamefont {Grant}, \citenamefont {Herskind}, \citenamefont
  {Dantan},\ and\ \citenamefont {Drewsen}}]{hendricks2007all}%
  \BibitemOpen
  \bibfield  {author} {\bibinfo {author} {\bibfnamefont {R.}~\bibnamefont
  {Hendricks}}, \bibinfo {author} {\bibfnamefont {D.}~\bibnamefont {Grant}},
  \bibinfo {author} {\bibfnamefont {P.~F.}\ \bibnamefont {Herskind}}, \bibinfo
  {author} {\bibfnamefont {A.}~\bibnamefont {Dantan}}, \ and\ \bibinfo {author}
  {\bibfnamefont {M.}~\bibnamefont {Drewsen}},\ }\href@noop {} {\bibfield
  {journal} {\bibinfo  {journal} {Applied Physics B}\ }\textbf {\bibinfo
  {volume} {88}},\ \bibinfo {pages} {507} (\bibinfo {year} {2007})}\BibitemShut
  {NoStop}%
\bibitem [{\citenamefont {Schnitzler}\ \emph {et~al.}(2010)\citenamefont
  {Schnitzler}, \citenamefont {Jacob}, \citenamefont {Fickler}, \citenamefont
  {Schmidt-Kaler},\ and\ \citenamefont {Singer}}]{schnitzler2010focusing}%
  \BibitemOpen
  \bibfield  {author} {\bibinfo {author} {\bibfnamefont {W.}~\bibnamefont
  {Schnitzler}}, \bibinfo {author} {\bibfnamefont {G.}~\bibnamefont {Jacob}},
  \bibinfo {author} {\bibfnamefont {R.}~\bibnamefont {Fickler}}, \bibinfo
  {author} {\bibfnamefont {F.}~\bibnamefont {Schmidt-Kaler}}, \ and\ \bibinfo
  {author} {\bibfnamefont {K.}~\bibnamefont {Singer}},\ }\href@noop {}
  {\bibfield  {journal} {\bibinfo  {journal} {New Journal of Physics}\ }\textbf
  {\bibinfo {volume} {12}},\ \bibinfo {pages} {065023} (\bibinfo {year}
  {2010})}\BibitemShut {NoStop}%
\bibitem [{\citenamefont {Pretko}\ and\ \citenamefont
  {Radzihovsky}(2018)}]{pretko2018fracton}%
  \BibitemOpen
  \bibfield  {author} {\bibinfo {author} {\bibfnamefont {M.}~\bibnamefont
  {Pretko}}\ and\ \bibinfo {author} {\bibfnamefont {L.}~\bibnamefont
  {Radzihovsky}},\ }\href@noop {} {\bibfield  {journal} {\bibinfo  {journal}
  {Physical Review Letters}\ }\textbf {\bibinfo {volume} {120}},\ \bibinfo
  {pages} {195301} (\bibinfo {year} {2018})}\BibitemShut {NoStop}%
\bibitem [{\citenamefont {Mielenz}\ \emph {et~al.}(2013)\citenamefont
  {Mielenz}, \citenamefont {Brox}, \citenamefont {Kahra}, \citenamefont
  {Leschhorn}, \citenamefont {Albert}, \citenamefont {Schaetz}, \citenamefont
  {Landa},\ and\ \citenamefont {Reznik}}]{mielenz2013trapping}%
  \BibitemOpen
  \bibfield  {author} {\bibinfo {author} {\bibfnamefont {M.}~\bibnamefont
  {Mielenz}}, \bibinfo {author} {\bibfnamefont {J.}~\bibnamefont {Brox}},
  \bibinfo {author} {\bibfnamefont {S.}~\bibnamefont {Kahra}}, \bibinfo
  {author} {\bibfnamefont {G.}~\bibnamefont {Leschhorn}}, \bibinfo {author}
  {\bibfnamefont {M.}~\bibnamefont {Albert}}, \bibinfo {author} {\bibfnamefont
  {T.}~\bibnamefont {Schaetz}}, \bibinfo {author} {\bibfnamefont
  {H.}~\bibnamefont {Landa}}, \ and\ \bibinfo {author} {\bibfnamefont
  {B.}~\bibnamefont {Reznik}},\ }\href@noop {} {\bibfield  {journal} {\bibinfo
  {journal} {Physical Review Letters}\ }\textbf {\bibinfo {volume} {110}},\
  \bibinfo {pages} {133004} (\bibinfo {year} {2013})}\BibitemShut {NoStop}%
\bibitem [{\citenamefont {Pyka}\ \emph {et~al.}(2013)\citenamefont {Pyka},
  \citenamefont {Keller}, \citenamefont {Partner}, \citenamefont {Nigmatullin},
  \citenamefont {Burgermeister}, \citenamefont {Meier}, \citenamefont
  {Kuhlmann}, \citenamefont {Retzker}, \citenamefont {Plenio}, \citenamefont
  {Zurek} \emph {et~al.}}]{pyka2013topological}%
  \BibitemOpen
  \bibfield  {author} {\bibinfo {author} {\bibfnamefont {K.}~\bibnamefont
  {Pyka}}, \bibinfo {author} {\bibfnamefont {J.}~\bibnamefont {Keller}},
  \bibinfo {author} {\bibfnamefont {H.}~\bibnamefont {Partner}}, \bibinfo
  {author} {\bibfnamefont {R.}~\bibnamefont {Nigmatullin}}, \bibinfo {author}
  {\bibfnamefont {T.}~\bibnamefont {Burgermeister}}, \bibinfo {author}
  {\bibfnamefont {D.}~\bibnamefont {Meier}}, \bibinfo {author} {\bibfnamefont
  {K.}~\bibnamefont {Kuhlmann}}, \bibinfo {author} {\bibfnamefont
  {A.}~\bibnamefont {Retzker}}, \bibinfo {author} {\bibfnamefont {M.~B.}\
  \bibnamefont {Plenio}}, \bibinfo {author} {\bibfnamefont {W.}~\bibnamefont
  {Zurek}},  \emph {et~al.},\ }\href@noop {} {\bibfield  {journal} {\bibinfo
  {journal} {Nature Communications}\ }\textbf {\bibinfo {volume} {4}},\
  \bibinfo {pages} {1} (\bibinfo {year} {2013})}\BibitemShut {NoStop}%
\bibitem [{\citenamefont {Brox}\ \emph {et~al.}(2017)\citenamefont {Brox},
  \citenamefont {Kiefer}, \citenamefont {Bujak}, \citenamefont {Schaetz},\ and\
  \citenamefont {Landa}}]{brox2017spectroscopy}%
  \BibitemOpen
  \bibfield  {author} {\bibinfo {author} {\bibfnamefont {J.}~\bibnamefont
  {Brox}}, \bibinfo {author} {\bibfnamefont {P.}~\bibnamefont {Kiefer}},
  \bibinfo {author} {\bibfnamefont {M.}~\bibnamefont {Bujak}}, \bibinfo
  {author} {\bibfnamefont {T.}~\bibnamefont {Schaetz}}, \ and\ \bibinfo
  {author} {\bibfnamefont {H.}~\bibnamefont {Landa}},\ }\href@noop {}
  {\bibfield  {journal} {\bibinfo  {journal} {Physical Review Letters}\
  }\textbf {\bibinfo {volume} {119}},\ \bibinfo {pages} {153602} (\bibinfo
  {year} {2017})}\BibitemShut {NoStop}%
\bibitem [{\citenamefont {Retzker}\ \emph {et~al.}(2008)\citenamefont
  {Retzker}, \citenamefont {Thompson}, \citenamefont {Segal},\ and\
  \citenamefont {Plenio}}]{retzker2008double}%
  \BibitemOpen
  \bibfield  {author} {\bibinfo {author} {\bibfnamefont {A.}~\bibnamefont
  {Retzker}}, \bibinfo {author} {\bibfnamefont {R.~C.}\ \bibnamefont
  {Thompson}}, \bibinfo {author} {\bibfnamefont {D.~M.}\ \bibnamefont {Segal}},
  \ and\ \bibinfo {author} {\bibfnamefont {M.~B.}\ \bibnamefont {Plenio}},\
  }\href@noop {} {\bibfield  {journal} {\bibinfo  {journal} {Physical Review
  Letters}\ }\textbf {\bibinfo {volume} {101}},\ \bibinfo {pages} {260504}
  (\bibinfo {year} {2008})}\BibitemShut {NoStop}%
\bibitem [{\citenamefont {Shimshoni}\ \emph {et~al.}(2011)\citenamefont
  {Shimshoni}, \citenamefont {Morigi},\ and\ \citenamefont
  {Fishman}}]{shimshoni2011quantum}%
  \BibitemOpen
  \bibfield  {author} {\bibinfo {author} {\bibfnamefont {E.}~\bibnamefont
  {Shimshoni}}, \bibinfo {author} {\bibfnamefont {G.}~\bibnamefont {Morigi}}, \
  and\ \bibinfo {author} {\bibfnamefont {S.}~\bibnamefont {Fishman}},\
  }\href@noop {} {\bibfield  {journal} {\bibinfo  {journal} {Physical Review
  Letters}\ }\textbf {\bibinfo {volume} {106}},\ \bibinfo {pages} {010401}
  (\bibinfo {year} {2011})}\BibitemShut {NoStop}%
\bibitem [{\citenamefont {H{\"a}rter}\ and\ \citenamefont
  {Hecker~Denschlag}(2014)}]{harter2014cold}%
  \BibitemOpen
  \bibfield  {author} {\bibinfo {author} {\bibfnamefont {A.}~\bibnamefont
  {H{\"a}rter}}\ and\ \bibinfo {author} {\bibfnamefont {J.}~\bibnamefont
  {Hecker~Denschlag}},\ }\href@noop {} {\bibfield  {journal} {\bibinfo
  {journal} {Contemporary Physics}\ }\textbf {\bibinfo {volume} {55}},\
  \bibinfo {pages} {33} (\bibinfo {year} {2014})}\BibitemShut {NoStop}%
\bibitem [{\citenamefont {Tomza}\ \emph {et~al.}(2019)\citenamefont {Tomza},
  \citenamefont {Jachymski}, \citenamefont {Gerritsma}, \citenamefont
  {Negretti}, \citenamefont {Calarco}, \citenamefont {Idziaszek},\ and\
  \citenamefont {Julienne}}]{tomza2019cold}%
  \BibitemOpen
  \bibfield  {author} {\bibinfo {author} {\bibfnamefont {M.}~\bibnamefont
  {Tomza}}, \bibinfo {author} {\bibfnamefont {K.}~\bibnamefont {Jachymski}},
  \bibinfo {author} {\bibfnamefont {R.}~\bibnamefont {Gerritsma}}, \bibinfo
  {author} {\bibfnamefont {A.}~\bibnamefont {Negretti}}, \bibinfo {author}
  {\bibfnamefont {T.}~\bibnamefont {Calarco}}, \bibinfo {author} {\bibfnamefont
  {Z.}~\bibnamefont {Idziaszek}}, \ and\ \bibinfo {author} {\bibfnamefont
  {P.~S.}\ \bibnamefont {Julienne}},\ }\href@noop {} {\bibfield  {journal}
  {\bibinfo  {journal} {Reviews of Modern Physics}\ }\textbf {\bibinfo {volume}
  {91}},\ \bibinfo {pages} {035001} (\bibinfo {year} {2019})}\BibitemShut
  {NoStop}%
\bibitem [{\citenamefont {Grier}\ \emph {et~al.}(2009)\citenamefont {Grier},
  \citenamefont {Cetina}, \citenamefont {Oru{\v{c}}evi{\'c}},\ and\
  \citenamefont {Vuleti{\'c}}}]{grier2009observation}%
  \BibitemOpen
  \bibfield  {author} {\bibinfo {author} {\bibfnamefont {A.~T.}\ \bibnamefont
  {Grier}}, \bibinfo {author} {\bibfnamefont {M.}~\bibnamefont {Cetina}},
  \bibinfo {author} {\bibfnamefont {F.}~\bibnamefont {Oru{\v{c}}evi{\'c}}}, \
  and\ \bibinfo {author} {\bibfnamefont {V.}~\bibnamefont {Vuleti{\'c}}},\
  }\href@noop {} {\bibfield  {journal} {\bibinfo  {journal} {Physical Review
  Letters}\ }\textbf {\bibinfo {volume} {102}},\ \bibinfo {pages} {223201}
  (\bibinfo {year} {2009})}\BibitemShut {NoStop}%
\bibitem [{\citenamefont {Hall}\ \emph {et~al.}(2013)\citenamefont {Hall},
  \citenamefont {Aymar}, \citenamefont {Raoult}, \citenamefont {Dulieu},\ and\
  \citenamefont {Willitsch}}]{hall2013light}%
  \BibitemOpen
  \bibfield  {author} {\bibinfo {author} {\bibfnamefont {F.~H.}\ \bibnamefont
  {Hall}}, \bibinfo {author} {\bibfnamefont {M.}~\bibnamefont {Aymar}},
  \bibinfo {author} {\bibfnamefont {M.}~\bibnamefont {Raoult}}, \bibinfo
  {author} {\bibfnamefont {O.}~\bibnamefont {Dulieu}}, \ and\ \bibinfo {author}
  {\bibfnamefont {S.}~\bibnamefont {Willitsch}},\ }\href@noop {} {\bibfield
  {journal} {\bibinfo  {journal} {Molecular Physics}\ }\textbf {\bibinfo
  {volume} {111}},\ \bibinfo {pages} {1683} (\bibinfo {year}
  {2013})}\BibitemShut {NoStop}%
\bibitem [{\citenamefont {Saito}\ \emph {et~al.}(2017)\citenamefont {Saito},
  \citenamefont {Haze}, \citenamefont {Sasakawa}, \citenamefont {Nakai},
  \citenamefont {Raoult}, \citenamefont {Da~Silva~Jr}, \citenamefont {Dulieu},\
  and\ \citenamefont {Mukaiyama}}]{saito2017characterization}%
  \BibitemOpen
  \bibfield  {author} {\bibinfo {author} {\bibfnamefont {R.}~\bibnamefont
  {Saito}}, \bibinfo {author} {\bibfnamefont {S.}~\bibnamefont {Haze}},
  \bibinfo {author} {\bibfnamefont {M.}~\bibnamefont {Sasakawa}}, \bibinfo
  {author} {\bibfnamefont {R.}~\bibnamefont {Nakai}}, \bibinfo {author}
  {\bibfnamefont {M.}~\bibnamefont {Raoult}}, \bibinfo {author} {\bibfnamefont
  {H.}~\bibnamefont {Da~Silva~Jr}}, \bibinfo {author} {\bibfnamefont
  {O.}~\bibnamefont {Dulieu}}, \ and\ \bibinfo {author} {\bibfnamefont
  {T.}~\bibnamefont {Mukaiyama}},\ }\href@noop {} {\bibfield  {journal}
  {\bibinfo  {journal} {Physical Review A}\ }\textbf {\bibinfo {volume} {95}},\
  \bibinfo {pages} {032709} (\bibinfo {year} {2017})}\BibitemShut {NoStop}%
\bibitem [{\citenamefont {Joger}\ \emph {et~al.}(2017)\citenamefont {Joger},
  \citenamefont {Furst}, \citenamefont {Ewald}, \citenamefont {Feldker},
  \citenamefont {Tomza},\ and\ \citenamefont
  {Gerritsma}}]{joger2017observation}%
  \BibitemOpen
  \bibfield  {author} {\bibinfo {author} {\bibfnamefont {J.}~\bibnamefont
  {Joger}}, \bibinfo {author} {\bibfnamefont {H.}~\bibnamefont {Furst}},
  \bibinfo {author} {\bibfnamefont {N.}~\bibnamefont {Ewald}}, \bibinfo
  {author} {\bibfnamefont {T.}~\bibnamefont {Feldker}}, \bibinfo {author}
  {\bibfnamefont {M.}~\bibnamefont {Tomza}}, \ and\ \bibinfo {author}
  {\bibfnamefont {R.}~\bibnamefont {Gerritsma}},\ }\href@noop {} {\bibfield
  {journal} {\bibinfo  {journal} {Physical Review A}\ }\textbf {\bibinfo
  {volume} {96}},\ \bibinfo {pages} {030703(R)} (\bibinfo {year}
  {2017})}\BibitemShut {NoStop}%
\bibitem [{\citenamefont {Ratschbacher}\ \emph {et~al.}(2013)\citenamefont
  {Ratschbacher}, \citenamefont {Sias}, \citenamefont {Carcagni}, \citenamefont
  {Silver}, \citenamefont {Zipkes},\ and\ \citenamefont
  {K{\"o}hl}}]{ratschbacher2013decoherence}%
  \BibitemOpen
  \bibfield  {author} {\bibinfo {author} {\bibfnamefont {L.}~\bibnamefont
  {Ratschbacher}}, \bibinfo {author} {\bibfnamefont {C.}~\bibnamefont {Sias}},
  \bibinfo {author} {\bibfnamefont {L.}~\bibnamefont {Carcagni}}, \bibinfo
  {author} {\bibfnamefont {J.~M.}\ \bibnamefont {Silver}}, \bibinfo {author}
  {\bibfnamefont {C.}~\bibnamefont {Zipkes}}, \ and\ \bibinfo {author}
  {\bibfnamefont {M.}~\bibnamefont {K{\"o}hl}},\ }\href@noop {} {\bibfield
  {journal} {\bibinfo  {journal} {Physical Review Letters}\ }\textbf {\bibinfo
  {volume} {110}},\ \bibinfo {pages} {160402} (\bibinfo {year}
  {2013})}\BibitemShut {NoStop}%
\bibitem [{\citenamefont {F{\"u}rst}\ \emph {et~al.}(2018)\citenamefont
  {F{\"u}rst}, \citenamefont {Feldker}, \citenamefont {Ewald}, \citenamefont
  {Joger}, \citenamefont {Tomza},\ and\ \citenamefont
  {Gerritsma}}]{furst2018dynamics}%
  \BibitemOpen
  \bibfield  {author} {\bibinfo {author} {\bibfnamefont {H.}~\bibnamefont
  {F{\"u}rst}}, \bibinfo {author} {\bibfnamefont {T.}~\bibnamefont {Feldker}},
  \bibinfo {author} {\bibfnamefont {N.~V.}\ \bibnamefont {Ewald}}, \bibinfo
  {author} {\bibfnamefont {J.}~\bibnamefont {Joger}}, \bibinfo {author}
  {\bibfnamefont {M.}~\bibnamefont {Tomza}}, \ and\ \bibinfo {author}
  {\bibfnamefont {R.}~\bibnamefont {Gerritsma}},\ }\href@noop {} {\bibfield
  {journal} {\bibinfo  {journal} {Physical Review A}\ }\textbf {\bibinfo
  {volume} {98}},\ \bibinfo {pages} {012713} (\bibinfo {year}
  {2018})}\BibitemShut {NoStop}%
\bibitem [{\citenamefont {Sikorsky}\ \emph {et~al.}(2018)\citenamefont
  {Sikorsky}, \citenamefont {Meir}, \citenamefont {Ben-Shlomi}, \citenamefont
  {Akerman},\ and\ \citenamefont {Ozeri}}]{sikorsky2018spin}%
  \BibitemOpen
  \bibfield  {author} {\bibinfo {author} {\bibfnamefont {T.}~\bibnamefont
  {Sikorsky}}, \bibinfo {author} {\bibfnamefont {Z.}~\bibnamefont {Meir}},
  \bibinfo {author} {\bibfnamefont {R.}~\bibnamefont {Ben-Shlomi}}, \bibinfo
  {author} {\bibfnamefont {N.}~\bibnamefont {Akerman}}, \ and\ \bibinfo
  {author} {\bibfnamefont {R.}~\bibnamefont {Ozeri}},\ }\href@noop {}
  {\bibfield  {journal} {\bibinfo  {journal} {Nature Communications}\ }\textbf
  {\bibinfo {volume} {9}},\ \bibinfo {pages} {1} (\bibinfo {year}
  {2018})}\BibitemShut {NoStop}%
\bibitem [{\citenamefont {Feldker}\ \emph {et~al.}(2020)\citenamefont
  {Feldker}, \citenamefont {F{\"u}rst}, \citenamefont {Hirzler}, \citenamefont
  {Ewald}, \citenamefont {Mazzanti}, \citenamefont {Wiater}, \citenamefont
  {Tomza},\ and\ \citenamefont {Gerritsma}}]{feldker2020buffer}%
  \BibitemOpen
  \bibfield  {author} {\bibinfo {author} {\bibfnamefont {T.}~\bibnamefont
  {Feldker}}, \bibinfo {author} {\bibfnamefont {H.}~\bibnamefont {F{\"u}rst}},
  \bibinfo {author} {\bibfnamefont {H.}~\bibnamefont {Hirzler}}, \bibinfo
  {author} {\bibfnamefont {N.}~\bibnamefont {Ewald}}, \bibinfo {author}
  {\bibfnamefont {M.}~\bibnamefont {Mazzanti}}, \bibinfo {author}
  {\bibfnamefont {D.}~\bibnamefont {Wiater}}, \bibinfo {author} {\bibfnamefont
  {M.}~\bibnamefont {Tomza}}, \ and\ \bibinfo {author} {\bibfnamefont
  {R.}~\bibnamefont {Gerritsma}},\ }\href@noop {} {\bibfield  {journal}
  {\bibinfo  {journal} {Nature Physics}\ ,\ \bibinfo {pages} {1}} (\bibinfo
  {year} {2020})}\BibitemShut {NoStop}%
\bibitem [{\citenamefont {Schmidt}\ \emph
  {et~al.}(2020{\natexlab{b}})\citenamefont {Schmidt}, \citenamefont
  {Weckesser}, \citenamefont {Thielemann}, \citenamefont {Schaetz},\ and\
  \citenamefont {Karpa}}]{schmidt2020optical}%
  \BibitemOpen
  \bibfield  {author} {\bibinfo {author} {\bibfnamefont {J.}~\bibnamefont
  {Schmidt}}, \bibinfo {author} {\bibfnamefont {P.}~\bibnamefont {Weckesser}},
  \bibinfo {author} {\bibfnamefont {F.}~\bibnamefont {Thielemann}}, \bibinfo
  {author} {\bibfnamefont {T.}~\bibnamefont {Schaetz}}, \ and\ \bibinfo
  {author} {\bibfnamefont {L.}~\bibnamefont {Karpa}},\ }\href@noop {}
  {\bibfield  {journal} {\bibinfo  {journal} {Physical Review Letters}\
  }\textbf {\bibinfo {volume} {124}},\ \bibinfo {pages} {053402} (\bibinfo
  {year} {2020}{\natexlab{b}})}\BibitemShut {NoStop}%
\end{thebibliography}%

\end{document}